\documentclass{svproc}
\usepackage{url}
\usepackage{amsmath}
\usepackage{xcolor}
\usepackage{subfig}
\usepackage{amssymb}

\usepackage{mwe}
\usepackage{graphbox}
\usepackage[utf8]{inputenc}
\usepackage{tabularx,ragged2e,booktabs,caption}
\newcommand{\rf}[1]{(\ref{#1})} 
\usepackage{rotating}
\usepackage{pgfplots}

\definecolor{o}{rgb}{0.8, 0.588, 0.204} 
\definecolor{db}{rgb}{0., 0.445, 0.738} 

\usepackage{tikz}
\newcommand{\norm}[1]{\left\lVert#1\right\rVert}
\newcommand{\edit}[1]{\textcolor{black}{#1}}
\DeclareMathOperator{\sign}{sign}

\begin{document}
\mainmatter              
\title{Surface Denoising based on Normal Filtering in a Robust Statistics Framework}
\titlerunning{Surface denoising}  
%
\author{Sunil Kumar Yadav \and Martin Skrodzki \and Eric Zimmermann \and Konrad Polthier}
\authorrunning{Yadav et al.} 
%
\tocauthor{Sunil Kumar Yadav, Martin Skrodzki, Eric Zimmermann, and Konrad Polthier}
\institute{Department of Computer Science and Mathematics, \\ Freie Universit\"at Berlin, Berlin, Germany\\
\email{sunil.yadav@fu-berlin.de}.\\ 
}

\maketitle              

\begin{abstract}
During a surface acquisition process using 3D scanners, noise is inevitable and an important step in geometry processing is to remove these noise components from these surfaces (given as points-set or triangulated mesh). The noise-removal process (denoising) can be performed by filtering the surface normals first and by adjusting the vertex positions according to filtered normals afterwards. Therefore, in many available denoising algorithms, the computation of noise-free normals is a key factor. A variety of filters have been introduced for noise-removal from normals, with different focus points like robustness against outliers or large amplitude of noise. Although these filters are performing well in different aspects, a unified framework is missing to establish the relation between them and to provide a theoretical analysis beyond the performance of each method.  

In this paper, we introduce such a framework to establish relations between a number of widely-used nonlinear filters for face normals in mesh denoising and vertex normals in point set denoising. 
We cover robust statistical estimation with M-smoothers and their application to linear and non-linear normal filtering.
Although these methods originate in different mathematical theories---which include diffusion-, bilateral-, and directional curvature-based algorithms---we demonstrate that all of them can be cast into a unified framework of robust statistics using robust error norms and their corresponding influence functions. 
This unification contributes to a better understanding of the individual methods and their relations with each other. 
Furthermore, the presented framework provides a platform for new techniques to combine the advantages of known filters and to compare them with available methods.
\end{abstract}

\section{Introduction}
\label{sec:Introdcution}
Surface denoising---generally being part of the preprocessing stage in the geometry processing pipeline---is designed to remove high-frequency noise corrupting a geometry. The noise generally arises from scanning or other acquisition processes. In contrast to smoothing, we are interested in preserving attributes and features of the geometry like edges and corners. Here, the difficulty lies in distinguishing these from noise, depending on the intensity of noise and the level of the attributes' details.

Denoising can therefore be considered as being part of the area of smoothing. It is used in all applications asking for a cleaned, i.e.\@ noise-free, surface with the additional property of keeping features. But more importantly, it is recognized as being a major tool in the preprocessing stage of geometry processing. The reason is that---besides computer designed models---the acquisition of real world models via 3D scanning processes unfortunately adds noise and outliers to the data due to mechanical limitations and sub-optimal surrounding conditions. These artifacts influence meshes and point sets alike and have to be removed to obtain a clean model for further use in different industry applications, e.g.\@ scientific analysis, automotive, medical diagnosis, rendering, and other geometry processing algorithms like surface reconstruction, feature detection, computer aided design, or 3D printing, see~\cite{Yadav2018Med} for applications in medical diagnoses and~\cite{2010Botsch_PolyMeshProc} for a variety of application scenarios.

A typical challenge arising in the denoising process is the decoupling of noise and features of a geometry. This is, because both are high-frequency components of the geometry in terms of the spectral setting. Other problems arise as noisy geometries include outliers, which are far away from the underlying ground truth. Furthermore, the amplitude of noise can be significant when compared to the feature size. To solve these problems, in both cases---for meshes and point sets---a variety of surface denoising algorithms have been published. These state-of-the-art methods can be categorized into:
\begin{enumerate}
	\item One-stage methods, where noise components are removed by adjusting the vertex positions based on the curvature information;
	\item Two-stage methods, where in the first stage, surface normals are filtered and then in the second stage vertex positions are adjusted according to the filtered normals. 
\end{enumerate} 
Two-stage methods are more effective in terms of feature-preservation as well as noise-removal and obtain minimum volume shrinkage compared to one-stage methods, see~\cite{Centin2018,Yadav2017,Yadav2018}. In the two-stage methods, surface normal filtering is the key part as it is responsible for both noise-removal and feature-preservation. Therefore, several procedures have been published for normal filtering. Each of these algorithms is effective in different aspects (like robustness against noise, feature preservation, detection of outliers, etc.). However, there is no unified theoretical framework available in which we can discuss the benefits and drawbacks of the normal filtering algorithms and in which we can derive the relations between these methods.

In this paper, we focus on this issue and introduce such a unified framework making use of robust statistics to derive relations between (both linear and non-linear) state-of-the-art surface normal filtering methods. On the basis of these relations, we discuss the robustness of each algorithm against noise and its respective feature-preservation capability. The presented framework can be used to provide pros and cons of published methods for the development of new algorithms. Furthermore, it can serve as a comparison possibility for such new procedures to state-of-the-art methods on a theoretically sound basis.

\subsection{Notation}
\label{sec:Notation}

Throughout the whole paper we will use the following notation. Let $ I, J, K $ denote index sets as subsets of~$ \mathbb{N} $. We consider a mesh~$ {\mathcal{M} = (P, E, F)} $ consisting of a set of points or vertices~$ {P = \lbrace p_i \rbrace_{i \in I} \subset \mathbb{R}^3} $ (which will be used in the point set setting as well), (undirected) edges~$ E $, and faces~$ F $. In general, we will assume that the mesh~$\mathcal{M}$ or the point set~$P$ is corrupted by noise. The set of normals is given as~$ {N = \lbrace n_j \rbrace_{j \in J} \subset \mathbb{S}^2} $, with~$ \mathbb{S}^2 $ the 2-dimensional unit-sphere in~$ \mathbb{R}^3 $ and neighborhoods are labeled~$ \Omega_k $ for~$ {k \in K} $. Sometimes we only refer to the neighborhood by~$ \Omega $ and to its representatives by~$ {p, q \in \Omega} $ without further labels, to simplify the notation where it is unambiguous. The used type of neighborhood will get specified when necessary and receive a dedicated index set, as it further depends on the context, i.e.\@ to which object (points, faces,~$ \ldots $) we are going to relate it. Consequently, normals and neighborhoods apply for faces and points depending whether we discuss the mesh or point set setting. Let~$ |X|$ denote the size of a set~$ X $ and let~\edit{$\norm{v}$} as well as $v^T$ be the Euclidean norm and the transpose of a vector~\edit{$ {v\in\mathbb{R}^3} $} respectively. \edit{A surface area or a vertex, both of high curvature (in comparison with the other elements of the geometry) will be referred to as a \emph{feature} of the mesh or the point set respectively.}

\subsection{Related Work}
\label{sec:RelatedWork}

In the last two decades, many surface smoothing algorithms have been developed. Due to the large number of available methods, for a comprehensive overview we refer to~\cite{2010Botsch_PolyMeshProc,Centin2018}. Here, we give a short overview of methods highly related to the robust statistics setting and of the most important state-of-the-art methods. 


As stated above, the removal of noise components is equivalent to the removal of high frequency components. Here, the Fourier transform is a common tool, allowing efficient implementations of low-pass filters to cut off high frequencies. It has been generalized to manifold harmonics to be applicable to 2-manifold surfaces via the eigenfunctions of the Laplace-Beltrami operator of these surfaces. Its matrix representation encodes the \emph{natural vibrations} of a triangle mesh in its eigenvectors and the \emph{natural frequencies} in its eigenvalues, see~\cite{1999Taubin_SignalProcFairSurfDes,2001Taubin_GeomSignProcMeshes}. One drawback is its cost for many applications as the eigenvector decomposition of the Laplace matrix is numerically challenging to compute, see~\cite{2008Vallet_SpecGeomProc}.

A similar removal of high-frequency components can be achieved by utilizing the diffusion flow, which dampens high frequencies (instead of cutting them off) by a multiplication with a Gaussian kernel. It can be computed directly on the mesh, making it cheaper and hence more practical than the Fourier transform. Let $ {f(\vec{p},t):\mathbb{R}^{3|P|+1}\rightarrow\mathbb{R}} $ be a given signal with $ {\vec{p} = (p_1, \ldots, p_{|P|})^{\edit{T}}} $. The diffusion equation:
\begin{equation}
\label{equ:DiffusionEquation}
	\frac{\partial f(\vec{p},t)}{\partial t} = \lambda \Delta f(\vec{p},t) 
\end{equation} 
describes the change of $ f $ over time by a scalar diffusion coefficient $ {\lambda\in\mathbb{R}} $ multiplied with its spatial Laplacian $ \Delta f $, which can be replaced by the Laplace-Beltrami operator on manifolds. As the discretization asks for small time steps to be numerically robust in the integration, the authors of~\cite{2001Desbrun_ImplFairingIrrMeshesCurvFlow} proposed an implicit time integration providing unconditional robustness even for large time steps. A smoothing procedure can be derived from this as update of the vertex positions~$ p_i $ by a point-wise update scheme
\begin{align}
\label{equ:DiffusionVertexUpdateScheme}
\begin{split}
				 & p_i \gets  p_i + h\lambda \Delta p_i,\\
	\text{with } & \Delta p_i = -2 H n_i,
\end{split}
\end{align}
because the Laplace-Beltrami operator on vertices corresponds to the mean curvature. Hence, all vertices $ p_i $ move in the corresponding normal direction $ n_i $ by a magnitude regulated by the mean curvature $ H $. This is known as the \emph{mean curvature flow}, see~\cite{2001Desbrun_ImplFairingIrrMeshesCurvFlow}. 

The isotropic Laplacian has been extended by a data-dependent diffusion tensor yielding the anisotropic flow equation:
\begin{equation}
\label{equ:AnisotropicDiffusionEquation}	
	\frac{\partial f}{\partial t} = \text{div}[ g_\sigma(\|\nabla f \|) \nabla f],
\end{equation} 
where $f$ is a signal as in Equation~\rf{equ:DiffusionEquation} and $g_\sigma(\cdot)$ is an edge stopping function (anisotropic weighting function), which is responsible for feature-preservation with a user input parameter $\sigma$ during denoising operations, see~\cite{1990Perona_scaleSpaceEdgeDetecAniDiff,2000Clarenz}. Further examples for the usage of the anisotropic diffusion equation can be found in~\cite{2003Bajaj_AnisoDiffSurf} and~\cite{2004Hildebrandt_AnisoFilteringSurfFeatures}. The same concept is extended to the context of point set smoothing by Lange and Polthier~\cite{LANGE2005} and to face normal filtering by Tasdizen et al.~\cite{Tasdizen2002}.

Another set of denoising techniques consists of two-stage mesh denoising algorithms. Here, at the first stage, face normals are filtered and in the second stage vertex positions are updated according to the newly computed face normals, see~\cite{Taubin2001}. Face normal filtering is performed by using several linear and non-linear filters in order to preserve sharp features~\cite{Centin2018,Yadav2017,Yagou2002,Yagou2003,Ohtake2002,Belyaev2001} and vertex updates are performed by using the edge-face orthogonality~\cite{Sun2007}.  

Finally, there are several denoising methods utilizing bilateral filtering. It arose from image processing~\cite{1998Tomasi_BilFilterGrayColImg} and uses a combination of two different weighting functions: a spatial kernel and a range kernel to preserve features and remove noise components. It got adapted to surface denoising for instance in~\cite{2003Fleishman_BilMeshDenoising}, where the information of spatial distances and the local variation of vertex normal vectors is combined for denoising. Bilateral filters are extended for face normal filtering, where a range kernel (Gaussian function) is defined based on the normal differences in the neighborhood~\cite{Yadav2018,Zheng2011}. A variation of bilateral filtering is also used extensively in mesh denoising in order to remove noise and retain sharp features~\cite{Jones2003,Zhang2015}.   

\subsection{Face Normal Filtering vs Vertex Position Filtering}
\label{sec:FaceNormalFilteringVSVertexPositionFiltering}

Broadly, surface smoothing algorithms can be divided into two categories, direct vertex position filtering, which is also known as one stage smoothing and two-stage filtering, which includes (face) normal filtering and vertex position updates as described above. 

\begin{figure}
	\centering
	\subfloat[Noisy\label{fig:ShrinkageComparison_Noisy}]{{\includegraphics[width=2.7cm]{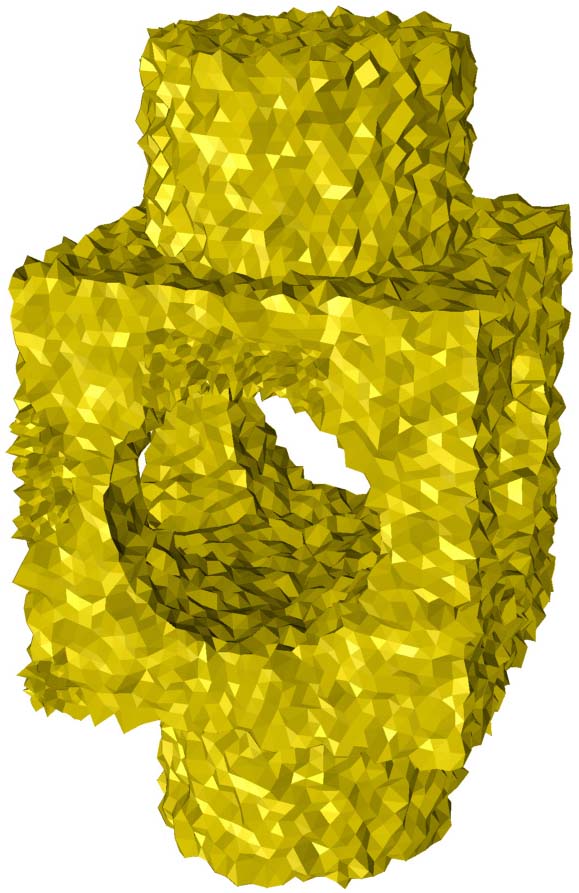} }}%
	\subfloat[Result~\cite{2004Hildebrandt_AnisoFilteringSurfFeatures}\label{fig:ShrinkageComparison_VertexNormal}]{{\includegraphics[width=2.8cm]{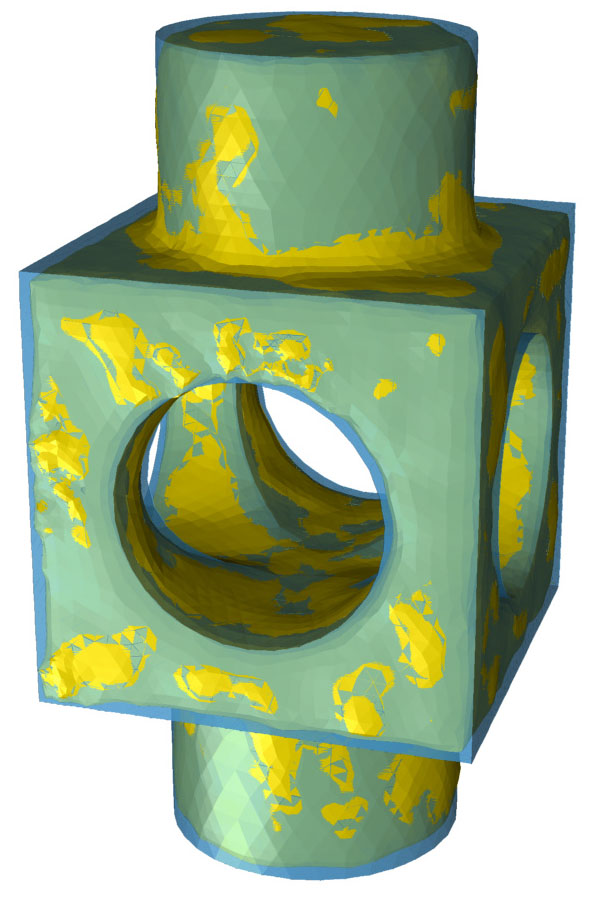} }}%
	\subfloat[Result~\cite{2003Fleishman_BilMeshDenoising}\label{fig:ShrinkageComparison_FaceNormal}]{{\includegraphics[width=2.7cm]{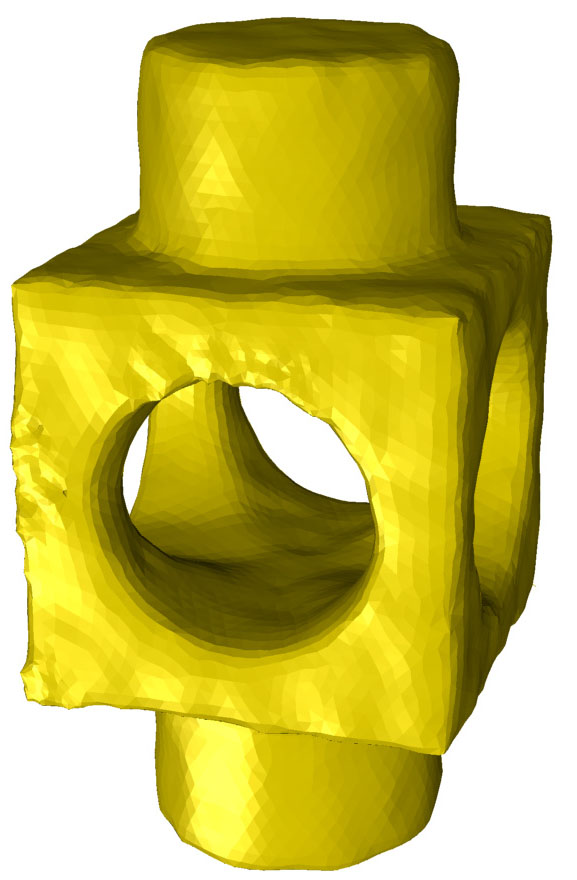} }}%
	\subfloat[Result~\cite{Yadav2017}\label{fig:ShrinkageComparison_FaceNormal2}]{{\includegraphics[width=2.65cm]{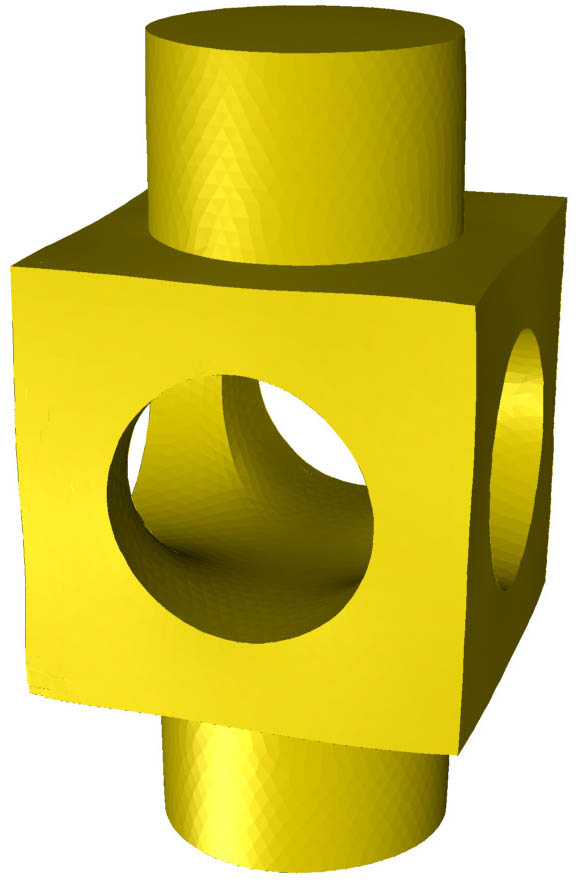} }}%
	\caption{A visual comparison between vertex position, vertex normal, and face normal filtering methods. Figure (a) shows the noisy block model, Figure (b) shows the denoised result of the method presented in~\cite{2004Hildebrandt_AnisoFilteringSurfFeatures}, based on mean curvature flow. More precisely, it shows two different surfaces, the original surface (green) and the denoised one (yellow). The difference between these two surfaces is visible due to volume shrinkage during the minimization. In contrast, Figures~(c) and (d) show the result of the face normal filtering methods~\cite{2003Fleishman_BilMeshDenoising} and~\cite{Yadav2017} respectively, which do not suffer from volume shrinkage.}
	\label{fig:comp}%
\end{figure}

Most of the one stage denoising algorithms (vertex position filtering) follow the concept of mean curvature flow, which is related to the Laplace-Beltrami operator and the mean curvature on the surface as shown in Equation~\rf{equ:DiffusionVertexUpdateScheme} and as discussed above. Basically, noise components are removed by minimizing the mean curvature on the surface, where the mean curvature is computed using the area gradient on the surface. Therefore, minimizing the curvature will result in minimizing the area, which will lead to volume shrinkage. This applies to most of the anisotropic and isotropic diffusion-based surface smoothing algorithms. These methods use vertex position filtering in their minimization. To illustrate this problem, Figure~\ref{fig:ShrinkageComparison_Noisy} shows a noisy model and Figure~\ref{fig:ShrinkageComparison_VertexNormal} shows the result obtained by using the mean curvature flow-based method of~\cite{2004Hildebrandt_AnisoFilteringSurfFeatures}. More precisely, Figure~\ref{fig:ShrinkageComparison_VertexNormal} shows two different surfaces, the original surface (green) and the denoised one (yellow). The difference between these two surfaces is visible due to volume shrinkage during the minimization.

On the other hand, in two-stage surface denoising, noise removal is performed based on the face normals. Basically, face normals are treated as signals on the vertices of the dual graph of the mesh with values in the unit sphere. The face normal denoising is generally performed by rotating the face normals on the unit sphere according to the weighted average of the corresponding neighbor face normals (see Equation~\rf{equ:L2ErrorAverageNormal} for a formalization). In other words, for noise removal, we operate in the dual space of the mesh and minimize the variation of face normals. This operation does not involve the curvature minimization on the vertex positions. Therefore, in two-stage surface denoising algorithms, volume shrinkage is minimal, as shown in Figures~\ref{fig:ShrinkageComparison_FaceNormal} and~\ref{fig:ShrinkageComparison_FaceNormal2}. 

Furthermore, in two-stage surface denoising, noise removal can be performed also on vertex normals~\cite{2003Fleishman_BilMeshDenoising} instead of face normals. However, in terms of sharp feature preservation, vertex normal filtering will not be as effective as face normal filtering because of the following reasons:
\begin{enumerate}
	\item The vertex normals of a mesh are usually derived from face normals. Therefore, processing face normals will avoid the ill-posedness and increase the robustness of the algorithm.
	\item At a sharp feature, the angle between vertex normals is smaller than the angle between the face normals. Therefore, face normals are more robust in feature-preservation compared to vertex normals.
\end{enumerate} 
As shown in Figures~\ref{fig:ShrinkageComparison_FaceNormal} and~\ref{fig:ShrinkageComparison_FaceNormal2}, face normal filtering better preserves sharp features compared to vertex normal filtering methods. However, in the context of point set surfaces, face normals are not available and denoising has to be performed using vertex normals.

\subsection{Scope}
\label{sec:Scope}

From our discussion in the last section, it is clear that the two-stage surface denoising algorithms are robust and efficient in terms of noise removal and feature-preservation. Therefore, in this article, we will cover surface normal filtering (face normal in the context of mesh surfaces and vertex normals in the context of point set surfaces) in a robust statistics framework. 

In the context of surface denoising, the most challenging task is to decouple sharp features from noise to treat them appropriately. Robust statistics is an efficient tool to identify the deviating substructures (\emph{outliers}) from the bulk data. Here, we will treat features on the geometry as \emph{outliers} because we want to deal with features differently compared to the non-feature areas. Based on this assumption, we derive relationships between different state-of-the-art methods for surface normal filtering using the concept of the robust error norm and its corresponding influence functions, see Section~\ref{sec:RobustStatisticalEsimation}. We also discuss the robustness of these algorithms within the presented framework, see Sections~\ref{sec:FaceNormalFilteringINTheRobustStatisticsFramework} and~\ref{sec:PointSetSurfaceDenoising}.


\section{Robust Statistical Estimation}
\label{sec:RobustStatisticalEsimation}

This article is concerned with robust statistics handling outliers during statistical data modeling. The field of robust statistics has developed methods to handle outliers in the data modeling process, see~\cite{Mrazek2006}. These methods describe the structure of best fitting the bulk of the data and identifying deviating substructures (outliers), see~\cite{Black1996}. In this section, we translate the robust statistics framework to the setting of surface denoising. As explained above, surface denoising is a preprocessing operation in many geometry processing algorithms, which removes noise components and retains sharp features. In the robust statistics framework, surface features can be seen as outliers and methods from robust statics can identify these, which in turn can be treated differently for feature-preserving surface denoising, see~\cite{Yadav2018}. As stated in the notation, we consider both a face and a vertex of the surface mesh to be a \emph{feature} respectively, if the corresponding normals of its neighbors have a high variation. Note that this is also the case for noisy faces and vertices, but not for outliers as they will not have a close neighborhood.

As reasoned in Section~\ref{sec:Scope}, we focus on two-stage mesh denoising algorithms. Recall that---as it is mentioned in Section~\ref{sec:Notation}---the surface $\mathcal{M}$ is corrupted by noise. Therefore, the vertices $ P $ and face normals $ N $ contain noise components, too. Let us first assume that the noise-free surface is represented by $\hat{\mathcal{M}}$ with $\hat{P}$ and $\hat{N}$ its vertices and face normals respectively. The noisy and noise-free face normals can be related by:
\begin{equation}
\label{equ:NoisyFromNoisefreeNormal}
	n = \hat{n} + \eta,
\end{equation}        
where $\eta$ is a random variable representing the noise corrupting the surface. If~$ \eta $ is a zero-mean Gaussian random variable and the surface is flat, then the denoised face normals can be computed by minimizing the following~$L_2$ error to compute the mean:
\begin{align}
\label{equ:L2ErrorAverageNormal}
	E(\hat{n}) = \sum_{n \in \Omega} \norm{\hat{n} - n}^2,&& \hat{n} = \frac{1}{\left|\Omega\right|}\sum_{n \in \Omega} n.
\end{align}
However, in real life scenarios, the noise $\eta$ is not always normally distributed and surfaces have sharp features, which can be seen as outliers. Therefore, in the following we will aim at computing an approximation~$\tilde{n}$ of~$\hat{n}$. To deal with this complicated situation, we use robust error norms, which lead to the theory of M-estimators, see Section~\ref{sec:M-Estimates} for details. An M-estimator of a face normal from noisy normals can be obtained as the minimum of the following error functional:
\begin{equation}
\label{equ:RhoEnergy}
	E_\sigma(\tilde{n}) = \sum_{n \in \Omega} \rho_\sigma \left (\norm{\tilde{n} - n}\right),
\end{equation}       
where $\rho_\sigma(\cdot):\mathbb{R}\rightarrow\mathbb{R}$ is a loss function and commonly called $\rho$-function or error norm~\cite{Black1996,Black1998,Durand2002} and the quantity~$\sigma$ is a user input. See Table~\ref{tab:estimators} for different choices for~$\rho_\sigma$. To minimize the effect of outliers, the loss function should not grow rapidly. To see the growing speed of the robust error norm $\rho_\sigma(\cdot)$, its derivative is computed, which is referred to as influence function ($\psi_\sigma(\cdot)$) in robust statics~\cite{Winkler1998}. Thus, the loss function and influence function are related as follows
\begin{equation}
\rho'_\sigma(x) =: \psi_\sigma(x),
\end{equation}
where for convenience, let us put $ x:=\norm{\tilde{n} - n}$.

During mesh denoising, at sharp features, the effect of the influence function should be minimal. The input parameter $x$ will be related to features, i.e.\@ to the variation of normals. Therefore, when~\edit{${x \rightarrow \infty}$}, the influence function should be zero, that is
\begin{equation*}
	\lim_{\edit{x\rightarrow\infty}}\psi_\sigma(x) = 0.
\end{equation*}
In our setting, feature values~($x$) are basically defined by the variation of normals, which is measured by the differences between the neighboring normals~$n_j$ and the central normal~$n_i$. However, these differences cannot approach infinity practically as~${n_i,n_j\in\mathbb{S}^2}$ for all~$i,j\in I$. Therefore, the above equation indicates that for bigger values of~$x$ the influence function should be diminished.    

Equation~\rf{equ:RhoEnergy} can be extended to take into account spatial weights in local neighborhoods using the following formulation:
\begin{equation}
\label{equ:RhoSpatial}
	E_{\sigma,\sigma_d}(\tilde{n}) = \sum_{n \in \Omega} \rho_\sigma\left( \norm{\tilde{n} - n}\right) f_{\sigma_d}(d),
\end{equation}
where the function $f_{\sigma_d}(d):\mathbb{R}\rightarrow\mathbb{R}$ is an isotropic \edit{weighting} factor, which takes the spatial distance $d$ between the considered geometry elements as the input argument and is responsible for smoothing out high frequency components of the geometry. The term~$\sigma_d$ controls the width of the spatial kernel and generally depends on the resolution (sampling density) of the given geometry. In case of mesh denoising, the distance is computed between the centroid of neighboring faces and the processed central face. For point set denoising, the term $d$ is computed between neighboring vertices and the processed central vertex. 

Throughout the whole paper, concerning the error functionals, we are going to ignore constant factors in the arguments for both the isotropic~($ \sigma_d $) and the anisotropic~($ \sigma $) case. This is to focus on the qualitative differences between the presented methods rather than on smaller variations.

\subsection{M-Estimators}
\label{sec:M-Estimates}

M-estimators are collections of different robust error norms to handle outliers. Any estimator defined by Equation~\rf{equ:RhoEnergy} is called an ``M-estimator''. The name comes from the generalized maximum likelihood concept, which can be deduced from Equation~\rf{equ:RhoEnergy}, when $-\rho_\sigma(x)$ is the likelihood function. Then, minimizing the energy $E_\sigma(\cdot)$ of Equation~\rf{equ:RhoEnergy} will be equivalent to the maximum likelihood estimate~\cite{Chu1998,Hampel2005}. As motivated above, in general, the robust estimators should have the following two properties:
\begin{enumerate}
	\item The error norm~$\rho_\sigma(x)$ should not grow rapidly.
	\item The influence function~$\psi_\sigma(x) = \rho_\sigma'(x)$ should be bounded.  
\end{enumerate}
For an efficient mesh denoising procedure, the influence function should be a \emph{re-descending function}, i.e.~${\psi_\sigma(x) \rightarrow 0}$ when~${\edit{x \rightarrow \infty}}$. In this case, the corresponding error norm $\rho_\sigma(x)$ is called \emph{re-descending influence error norm}~\cite{Hampel2005}.   

In general, surface normal (i.e.\@ face and vertex normal) filtering is performed by computing weighted averages of neighboring normals, see Equation~\rf{equ:DiscreteFaceNormalAverage}. The weighting functions are vital for feature-preserving normal filtering and they can be either linear or non-linear. Here, we will formulate the relationship between weighting function, robust error norm, and the corresponding influence function.  
 
From Equation~\rf{equ:AnisotropicDiffusionEquation}, we know that the anisotropic diffusion is controlled by an edge stopping function, which is represented by $g_\sigma(x)$. In this article, we termed it as anisotropic weighting function. Equation~\rf{equ:RhoEnergy} can be minimized using gradient descent to update the surface normal:
\begin{equation}
\label{equ:RhoGradientDescent}
	\edit{n^{t+1} = n^t + \lambda \nabla E_\sigma(x) = n^t + \lambda \sum_{n \in \Omega}\nabla \rho_\sigma(\left\|\tilde{n}-n\right\|)},
\end{equation}
\edit{where $t$ is the iteration number and $\lambda$ represents the step size. Here,~$\rho_\sigma$ is interpreted as a concatenation, taking the norm of a vector as argument, while the norm receives~${(\tilde{n})\in\mathbb{R}^3}$ as argument. The complete function then maps from~$\mathbb{R}^3$ to~$\mathbb{R}$. The differentiation let us consider the gradient of~$\rho_{\sigma}$ as a natural generalization of the derivative in the one-dimensional case. Following the reasoning of~\cite{Jones2003}, also adapted by~\cite{Zheng2011}, we adapt the procedure introduced in~\cite{1998Tomasi_BilFilterGrayColImg} for signal processing to the context of mesh processing by feeding the normal distance~$x$---as defined above---into the error norm~$\rho_\sigma$ and a spatial distance into the spatial weighting function~$f_\sigma$. This analogy motivates us to analyze the following well-established relation from signal processing (consider for a specific derivation~\cite[Sections 4.1 and 5.3]{Black1996} and more generally~\cite{Hampel2005,Huber1981})},
\begin{equation}
\label{equ:RhoAndGRelation}
	g_\sigma(x) = \frac{\rho_\sigma'(x)}{x} =: \frac{\psi_\sigma(x)}{x}.
\end{equation}
\edit{Applications of this relation in image and geometry processing can be found in~\cite{Jones2003,Black1998,Durand2002}.}

The weighting function~$g_\sigma(x)$ should capture the anisotropic behavior of the mesh or the point set respectively and should be chosen based on the above relations in the robust statistics framework. Table~\ref{tab:estimators} consists of several well known M-estimators with their robust error norms, their influence functions, and their corresponding anisotropic weighting functions.

Equation~(\ref{equ:L2ErrorAverageNormal}) shows an example of an estimator with a quadratic error norm (${\rho_\sigma(x) = x^2}$). This norm grows rapidly and its influence function (${\psi_\sigma(x)=2x}$) is unbounded (non re-descending) as shown in Table~\ref{tab:estimators}. Therefore, the quadratic estimator is very sensitive to outliers and not useful in feature-preserving mesh denoising.
    
\begin{table}
	\pgfplotsset{		
		samples=150,
		axis lines=left,
		ticks=none,
		y axis line style=-,
		x axis line style=-,
	}
	\centering
	\captionof{table}{M-Estimators} \label{tab:estimators} 
	\begin{tabular}{m{5.cm}|m{1.7cm}|m{1.7cm}|m{1.7cm}}
		Error norm $\rho_\sigma(x)$& \edit{Error norm $\rho_\sigma(x)$} & \edit{Influence function ${\psi_\sigma(x)=\rho'_\sigma(x)}$} & \edit{Weighting function ${g_\sigma(x)=\frac{\psi_\sigma(x)}{x}}$} \\
		\hline
		$ \diamond $ $L_2$-norm \cite{Black1998}, independent of~$\sigma$,
		\begin{equation*}
			  \rho_\sigma(x) = x^2  
		\end{equation*}
		& \begin{tikzpicture}
		\begin{axis}[
		domain=0:2, xmax=2.1,
		restrict y to domain=0:4,
		y=0.25cm,
		x=0.5cm,
		]
		\addplot [db, thick] {x^2};
		\end{axis}
		\end{tikzpicture} 
		& \begin{tikzpicture}
		\begin{axis}[
		domain=0:2, xmax=2.1,
		restrict y to domain=0:2,
		y=0.5cm,
		x=0.5cm,
		]
		\addplot [db, thick] {x};
		\end{axis}
		\end{tikzpicture}
		& \begin{tikzpicture}
		\begin{axis}[
		domain=0:2, xmax=2.1,
		restrict y to domain=0:2,
		ymax=2,
		ymin=0,
		y=0.5cm,
		x=0.5cm,
		]
		\addplot [db, thick] {1};
		\end{axis}
		\end{tikzpicture} \\
		$ \diamond $ Truncated $ L_2 $-norm \cite{Black1996}
		\begin{equation*}
			\rho_\sigma(x) =
			\begin{cases}
				x^2 & |x| < \sqrt{\sigma}\\
				\sigma & otrw.
			\end{cases}
		\end{equation*}
		& \begin{tikzpicture}
		\begin{axis}[
		domain=0:2, xmax=2.1,
		restrict y to domain=0:2,
		ymax=2,
		y=0.5cm,
		x=0.5cm,
		]
		\addplot [db, domain=0:1, thick] {x^2};
		\addplot [db, domain=1:2, thick] {1};
		\end{axis}
		\end{tikzpicture}
		& \begin{tikzpicture}
		\begin{axis}[
		domain=0:2, xmax=2.1,
		restrict y to domain=0:2,
		ymax=2,
		y=0.5cm,
		x=0.5cm,
		]
		\addplot [db, domain=0:1, thick] {x};
		\addplot [db, domain=1:2, very thick] {0};
		\end{axis}
		\end{tikzpicture}
		& \begin{tikzpicture}
		\begin{axis}[
		domain=0:2, xmax=2.1,
		restrict y to domain=0:2,
		ymax=2,
		y=0.5cm,
		x=0.5cm,
		]
		\addplot [db, domain=0:1, thick] {1};
		\addplot [db, domain=1:2, very thick] {0};
		\end{axis}
		\end{tikzpicture}\\
		$ \diamond $ $L_1$-norm \cite{Hampel2005}, independent of~$\sigma$,
		\begin{equation*}
			\rho_\sigma(x) = |x|
		\end{equation*} 
		& \begin{tikzpicture}
		\begin{axis}[
		domain=0:2, xmax=2.1,
		restrict y to domain=0:2,
		y=0.5cm,
		x=0.5cm,
		]
		\addplot [db, thick] {x};
		\end{axis}
		\draw[db, fill=white] (0,0) circle (2pt);
		\end{tikzpicture}
		& \begin{tikzpicture}
		\begin{axis}[
		domain=0:2, xmax=2.1,
		restrict y to domain=0:2,
		y=2.5cm,
		x=0.5cm,
		]
		\addplot [db, thick] {1};
		\end{axis}
		\draw[db, fill=white] (0,0.5) circle (2pt);
		\end{tikzpicture}
		&\begin{tikzpicture}
		\begin{axis}[
		domain=0:2, xmin=0, xmax=2.1,
		restrict y to domain=0:2,
		ymax=2,
		ymin=0,
		y=0.5cm,
		x=0.5cm,
		]
		\addplot [db, thick] {1/x};
		\end{axis}
		\end{tikzpicture}\\
		$ \diamond $ Truncated $L_1$-norm \cite{Hampel2005}
		\begin{equation*}
			\rho_\sigma(x) =
			\begin{cases}
				|x| & |x| < \sigma\\
				\sigma & otrw.
			\end{cases}
		\end{equation*}		
		& \begin{tikzpicture}
		\begin{axis}[
		domain=0:2, xmax=2.1,
		restrict y to domain=0:2,
		ymax=2,
		y=0.5cm,
		x=0.5cm,
		]
		\addplot [db, domain=0:1, thick] {x};
		\addplot [db, domain=1:2, thick] {1};
		\end{axis}
		\end{tikzpicture}
		& \begin{tikzpicture}
		\begin{axis}[
		domain=0:2, xmax=2.1,
		restrict y to domain=0:2,
		ymax=2,
		y=0.5cm,
		x=0.5cm,
		]
		\addplot [db, domain=0:1, thick] {1};
		\addplot [db, domain=1:2, very thick] {0};
		\end{axis}
		\end{tikzpicture}
		& \begin{tikzpicture}
		\begin{axis}[
		domain=0:2, xmin=0, xmax=2.1,
		restrict y to domain=0:2,
		ymax=2,
		y=0.5cm,
		x=0.5cm,
		]
		\addplot [db, domain=0:1, thick] {1/x};
		\addplot [db, domain=1:2, very thick] {0};
		\end{axis}
		\end{tikzpicture}\\
		$ \diamond $ Huber's minimax \cite{Huber1981}
		\begin{equation*}
		\rho_\sigma(x) =
			\begin{cases}
				\frac{x^2}{2 \sigma} + \frac{\sigma}{2} & |x| < \sigma\\
				|x| & otrw.
			\end{cases}
		\end{equation*}		
		& \begin{tikzpicture}
		\begin{axis}[
		domain=0:2, xmax=2.1,
		restrict y to domain=0:2,
		ymax=2,
		ymin=0,
		y=0.5cm,
		x=0.5cm,
		]
		\addplot [db, domain=0:0.25, thick] {(2*x^2) + 1/8};
		\addplot [db, domain=0.25:2, thick] {x};
		\end{axis}
		\end{tikzpicture} 
		& \begin{tikzpicture}
		\begin{axis}[
		domain=0:2, xmax=2.1,
		restrict y to domain=0:2,
		ymax=2,
		ymin=0,
		y=0.5cm,
		x=0.5cm,
		]
		\addplot [db, domain=0:0.25, thick] {4*x};
		\addplot [db, domain=0.25:2, thick] {1};
		\end{axis}
		\end{tikzpicture} 
		& \begin{tikzpicture}
		\begin{axis}[
		domain=0:2, xmax=2.1,
		restrict y to domain=0:5,
		ymax=5,
		ymin=0,
		y=0.2cm,
		x=0.5cm,
		]
		\addplot [db, domain=0:0.25, thick] {4};
		\addplot [db, domain=0.25:2, thick] {1/x};
		\end{axis}
		\end{tikzpicture}\\
		$ \diamond $ Lorentzian-norm \cite{Black1998}
		\begin{equation*}
			\rho_\sigma(x) = \log\left[1+\frac{1}{2}\left(\frac{x}{\sigma}\right)^2\right]
		\end{equation*} 
		& \begin{tikzpicture}
		\begin{axis}[
		domain=0:10, xmax=10,
		restrict y to domain=0:5,
		ymax=5,
		ymin=0,
		y=0.2cm,
		x=0.1cm,
		]
		\addplot [db, domain=0:10, thick] {ln(1+0.5*(x/1)^2)};
		\end{axis}
		\end{tikzpicture}
		& \begin{tikzpicture}
		\begin{axis}[
		domain=0:10, xmax=10,
		restrict y to domain=0:1,
		ymax=1,
		ymin=0,
		y=1cm,
		x=0.1cm,
		]
		\addplot [db, domain=0:10, thick] {1/(1+0.5*(x/1)^2)*(x)};
		\end{axis}
		\end{tikzpicture}
		& \begin{tikzpicture}
		\begin{axis}[
		domain=0:10, xmax=10,
		restrict y to domain=0:1,
		ymax=1,
		ymin=0,
		y=1cm,
		x=0.1cm,
		]
		\addplot [db, domain=0:10, thick] {1/(1+0.5*(x/1)^2)};
		\end{axis}
		\end{tikzpicture}\\
		$ \diamond $ Gaussian norm \cite{Black1996}
		\begin{equation*}
			\rho_\sigma(x) = 1 - e^{\left(-\frac{x^2}{\sigma^2}\right)}
		\end{equation*} 
		& \begin{tikzpicture}
		\begin{axis}[
		domain=0:4, xmax=4,
		restrict y to domain=0:1.1,
		ymax=1.1,
		ymin=0,
		y=1cm,
		x=0.25cm,
		]
		\addplot [db, domain=0:4, thick] {1-e^(-x^2)};
		\end{axis}
		\end{tikzpicture}
		& \begin{tikzpicture}
		\begin{axis}[
		domain=0:4, xmax=4,
		restrict y to domain=0:1.1,
		ymax=1.1,
		ymin=0,
		y=1cm,
		x=0.25cm,
		]
		\addplot [db, domain=0:4, thick] {2*x*e^(-x^2)};
		\end{axis}
		\end{tikzpicture}
		& \begin{tikzpicture}
		\begin{axis}[
		domain=0:3, xmax=3,
		restrict y to domain=0:2.1,
		ymax=2.1,
		ymin=0,
		y=0.5cm,
		x=0.33cm,
		]
		\addplot [db, domain=0:4, thick] {2*e^(-x^2)};
		\end{axis}
		\end{tikzpicture}\\
		$ \diamond $ Tukey's norm \cite{Beaton1974} 
		\begin{equation*}
		\rho_\sigma(x) =
			\begin{cases}
				\edit{\frac{x^2}{\sigma^2} - \frac{x^4}{\sigma^4} + \frac{x^6}{3 \sigma^6}} & |x| < \sigma\\
				\frac{1}{3} & otrw.
			\end{cases}
		\end{equation*}	
		& \begin{tikzpicture}
		\begin{axis}[
		domain=0:2, xmax=2.1,
		restrict y to domain=0:0.5,
		ymax=0.5,
		ymin=0,
		y=2cm,
		x=0.5cm,
		]
		\addplot [db, domain=0:1, thick] {x^2-x^4+1/3*x^6};
		\addplot [db, domain=1:2, thick] {1/3};
		\end{axis}
		\end{tikzpicture} 
		& \begin{tikzpicture}
		\begin{axis}[
		domain=0:2, xmax=2.1,
		restrict y to domain=0:0.75,
		ymax=0.75,
		ymin=0,
		y=1.33cm,
		x=0.5cm,
		]
		\addplot [db, domain=0:1, thick] {2*x-4*x^3+2*x^5};
		\addplot [db, domain=1:2, thick] {0};
		\end{axis}
		\end{tikzpicture}
		& \begin{tikzpicture}
		\begin{axis}[
		domain=0:2, xmax=2.1,
		restrict y to domain=0:2.1,
		ymax=2.1,
		ymin=0,
		y=0.5cm,
		x=0.5cm,
		]
		\addplot [db, domain=0:1, thick] {2-4*x^2+2*x^4};
		\addplot [db, domain=1:2, thick] {0};
		\end{axis}
		\end{tikzpicture}\\
		\hline
	\end{tabular}
\end{table}

The quadratic error norm can be truncated in order to convert it into a re-descending influence error norm. The second row of Table~\ref{tab:estimators} shows the truncated quadratic error norm that has a re-descending influence function~$\psi_\sigma(x)$ with a bounded error norm~$\rho_\sigma(x)$. However, the behavior of~$\psi_\sigma(x)$ is linearly increasing within the range of the user input~$\sigma$, which is not desired for feature preservation.

As shown in Table~\ref{tab:estimators}, the~$L_1$ error norm (${\rho_\sigma(x) = |x|}$, third row) and Huber's minimax error norm (fifth row) do not have re-descending influence functions even though they are bounded by a non-zero constant value. These two perform better in terms of separating outliers compared to the (truncated) quadratic error norm.

The other error norms listed in Table~\ref{tab:estimators}, which include the truncated $L_1$~error norm as well as the Lorentzian, Gaussian, and Tukey's norm have re-descending influence functions. Among all re-descending influence error norms, the truncated~$L_1$ and Tukey's error norm cut off the influence function's response strictly while the other norms have a non-zero influence function on a larger interval.


\section{Face Normal Filtering in the Robust Statistics Framework}
\label{sec:FaceNormalFilteringINTheRobustStatisticsFramework}

In this section, we will discuss state-of-the-art methods for face normal filtering utilizing the robust statistics framework and M-estimators as described above. Based on the relationship between the robust error norm, the influence function, and the weighting function as established in Equation~\rf{equ:RhoAndGRelation}, we will discuss the robustness and effectiveness of state-of-the-art methods for removing noise and preserving features.
 
The face normals $N$ of a triangulated mesh $\mathcal{M}$ can be seen as graph signals on the graph induced by the dual mesh of $\mathcal{M}$ with values in the unit sphere. The centroid of each face $f_i$ is denoted by $c_i$, which can be treated as the vertex position on the dual mesh. In general, the filtered face normal $\tilde{n}_i$ corresponding to a noisy face normal $n_i$ can be computed using the following equation:
\begin{equation}
\label{equ:DiscreteFaceNormalAverage}
\edit{\tilde{n}_i = \frac{1}{\omega} \sum_{j\in \Omega_i} g_\sigma\left (\norm{n_i - n_j}^2\right) f_{\sigma_d}(\norm{c_i -c_j}^2) n_j,}
\end{equation}
where \edit{$\omega = \left\|\sum_{j\in \Omega_i} g_\sigma(\norm{n_i - n_j}^2) f_{\sigma_d}(\norm{c_i -c_j}^2)n_j\right\|$} ensures~$\tilde{n}_i$ to be of unit-length. The term $ \Omega_i $ represents the mesh neighborhood around the $i$th triangle, which can be combinatorial or a geometrical disk of some (user-defined) radius. The above equation represents a general formula for face normal filtering and follows the error functional presented in Equation~\rf{equ:RhoSpatial}. The efficiency of this approach heavily depends on the choice of the weighting functions~$g_\sigma(\cdot)$ and~$f_{\sigma_d}(\cdot)$. 

\edit{In the following, we will present several state-of-the-art approaches for these choices. The listed algorithms use different input arguments for the robust error functionals. Common choices are the Euclidean distance of normals~$\left\|n_i-n_j\right\|$, the angle between two normals~$\angle(n_i,n_j)$, or the quantity~$\arccos(n_i\cdot n_j)$. We will stick to the notation used in the respective original paper in the following discussion. However, note that these input arguments are related. In particular, we obtain
\begin{align*}
\cos(\angle(n_i,n_j)) = \frac{n_i\cdot n_j}{\left\|n_i\right\|\left\|n_j\right\|}=n_i\cdot n_j \Rightarrow \angle(n_i,n_j)=\arccos(n_i\cdot n_j).
\end{align*} 
by the Euclidean scalar product because all normals considered are of unit-length. Furthermore, (by the law of cosines) it is
\begin{align*}
\left\|n_i-n_j\right\|^2& =\left\|n_i\right\|^2+\left\|n_j\right\|^2-2\cdot \left\|n_i\right\|\cdot\left\|n_j\right\|\cdot\cos(\angle(n_i,n_j))\\
	&= 2-2\cos(\angle(n_i,n_j))\\
\Rightarrow \angle(n_i,n_j) &= \arccos\left(1-\frac{\left\|n_i-n_j\right\|^2}{2}\right).
\end{align*}}

\subsection{\edit{Unilateral normal filtering}}
\label{sec:UnilateralNormalFiltering}

Unilateral normal filtering performs noise-removal from noisy normals using a single anisotropic kernel function. From our setup in Equation~\rf{equ:RhoSpatial}, it is clear that the unilateral normal filtering algorithms are using~$g_\sigma(x)$ as anisotropic weighting function while the spatial filter will be equal to one, i.e.~${f_{\sigma_d}(d)\equiv 1}$. These methods are effective against low intensity of noise and enhance sharp features. However, they are not robust against moderate or high levels of noise because of the unavailability of the spatial filter~$f_{\sigma_d}(d)$.

\subsubsection{\edit{a) Belyaev and Ohtake \cite{Belyaev2001}}}
 introduce non-linear diffusion of face normals to enhance the features of the geometry. Their algorithm uses the following weighting function:
\begin{equation}
\label{equ:belayev2002}
	g_\sigma(x) = \exp\left(-\frac{x^2}{\sigma^2}\right).
\end{equation}
This weight is a non-linear function and the input argument is encoding the directional curvature. It is given as
\begin{equation*}
	x = \frac{\angle(n_i, n_j)}{d},
\end{equation*}
\edit{where~$\angle(n_i, n_j)$ denotes the angle between~$n_i$ and~$n_j$}, the term~${d=\norm{c_i-c_j}}$ represents the distance between the centroids (as presented above) of the central face and its neighboring face, and~${n_i, n_j \in N}$ are face normals of the central face and its neighboring face, respectively. The term~$\sigma$ is a user input to better adapt the algorithm to the given geometry. It is chosen based on the amount of noise, curvature, and the resolution of the geometry. The directional curvature~$x$ measures the similarity between neighboring normals. In the robust statistics framework, by using Equation~\rf{equ:RhoAndGRelation}, we can deduce the used error norm as
\begin{equation}
	\label{equ:belayevRho}
	\rho_\sigma(x) = \int\limits_{0}^{x} x' g_\sigma(x') dx'= \frac{\sigma^2}{2} \left ( 1 - \exp\left(-\frac{x^2}{\sigma^2}\right) \right).
\end{equation}
Similarly, the influence function can be derived as
\begin{align}
	\label{equ:belayevPsi}
	\psi_\sigma(x) = x  g_\sigma(x) = x \exp\left(-\frac{x^2}{\sigma^2}\right), && 	\lim_{\edit{x\rightarrow\infty}}\psi_\sigma(x) = 0. 
\end{align}
The above two equations indicate that this algorithm applies the Gaussian error norm (second last row of Table~\ref{tab:estimators}), which has a re-descending influence function and makes the algorithm robust against outliers. However, the spatial smoothing function~$f_{\sigma_d}(\cdot)$ is not used in this algorithm, which reduces the robustness of the algorithm against significant noise.

\subsubsection{\edit{b) Yagou et al.~\cite{Yagou2002}}}
apply mean and median filtering to face normals. Mean filtering of normals is performed by simply uniformly averaging neighboring normals. Therefore, the anisotropic weighting function~${g_\sigma(x) \equiv 1}$ leads to an error norm and influence function of
\begin{align}
	\rho_\sigma(x) = \int_{0}^{x} x' g_\sigma(x') dx' = x^2 && \text{ and } && \psi_\sigma(x) = x g_\sigma(x)= x
\end{align} 
respectively. From the equation above, it is clear that mean filtering follows the quadratic error norm~(${\rho_\sigma(x)=x^2, g_\sigma(x) = 1}$) (the first row in Table~\ref{tab:estimators}) and it has an unbounded influence function ${(\lim_{\edit{x\rightarrow\infty}}\psi_\sigma(x) = \infty)}$, which makes the algorithm sensitive to outliers and produces feature blurring. \edit{This method uses the triangle area as a weighting function, i.e.\@ in the notation of Equation~\rf{equ:RhoSpatial}, it computes~$f_{\sigma_d}(d)$ for a given face~$f_i$ as~$\text{area}(f_i)$. However, this makes the algorithm only insensitive to irregular sampling.}

On the other hand, median filtering is estimated using the $L_1$~error norm~\cite{Hampel2005}. Therefore, the corresponding error norm and influence function can be derived as
\begin{align}
\rho_\sigma(x)=\edit{|x|} && \text{ and } && \psi_\sigma(x) = \rho'_\sigma(x)= \begin{cases}
1 & \edit{|x|} \neq 0\\
\text{undefined} & x=0.
\end{cases}
\end{align}
By using the relation from Equation~\rf{equ:RhoAndGRelation}, the anisotropic weighting function can be written as
\begin{equation}
	g_\sigma(x) = \frac{\psi_\sigma(x)}{x} = \begin{cases}
	\edit{\frac{1}{|x|}} & \edit{|x| \neq 0}\\
	\text{undefined} & x=0.
	\end{cases}
\end{equation}
In this algorithm, the input~$x$ is given by the Euclidean distance of the neighboring normal~${n_j \in N}$ to the central normal~$n_i$, i.e.~${x=\left\|n_i-n_j\right\|}$. The~$L_1$-norm is better compared to the quadratic error norm in terms of robustness to outliers. However, the corresponding influence function is not re-descending (see Table~\ref{tab:estimators}) and produces a constant value for outliers. 
 
Weighted median filtering is applying a spatial weighting function to provide higher weights to closer points compared to distant points, see~\cite{Yagou2002}. This weighting function is truncating the effect of local neighboring faces. Therefore, the weighted median follows a truncated $L_1$-norm and its corresponding influence function can be derived as
\begin{equation}
	\psi_\sigma(x) = \rho'_\sigma(x)= \begin{cases}
		0 & |x| < \sigma\\
		\sign(x) & 0 < \edit{|x| \leq \sigma}\\
		\text{undefined} & x=0 \\
	\end{cases}.
\end{equation}
By using the relation from Equation~\rf{equ:RhoAndGRelation}, the anisotropic weighting function can be written as
\begin{equation}
	g_\sigma(x) = \frac{\psi_\sigma(x)}{x} = \begin{cases}
	0 & |x| < \sigma\\
	\edit{\frac{\sign(x)}{x}} & \edit{0 < |x| \leq \sigma}\\
	\text{undefined} & x=0 \\
	\end{cases}.
\end{equation}
The truncated $L_1$-norm has a re-descending influence function, which enhances the feature preservation capability of the algorithm compared to mean and median filtering.
 
From the influence functions of the $L_1$-norm and the truncated $L_1$-norm, it is clear that these norms are capable of feature preservation during the process of face normal filtering. However, these influence functions and their corresponding anisotropic weighting functions are not well defined at~${x=0}$, which is not desirable. 

\subsubsection{\edit{c) Huber~\cite{Huber1981}}} \edit{proposes a slight modification of the weighting function before mentioned to overcome the issue of not being well-defined at $ {x = 0} $. He suggests}
\begin{align}
	\rho_\sigma(x)=\begin{cases}
		\frac{x^2}{2\sigma}+\frac{\sigma}{2} & |x|<\sigma\\
		|x| & \text{otrw}.
	\end{cases}.
\end{align}
This modified error norm is commonly known as Huber's minimax norm (see fifth row in Table~\ref{tab:estimators}). The corresponding influence and anisotropic weighting functions can be derived as
\begin{align}
	\psi_\sigma(x) = \begin{cases}
	\frac{x}{\sigma} & \edit{|x| < \sigma}\\
	\edit{\sign(x)} & \text{otrw}.
	\end{cases},  && g_\sigma(x) = \begin{cases}
	\frac{1}{\sigma} & \edit{|x| < \sigma}\\
	\frac{ \edit{\sign(x)} }{x}& \text{otrw}.
	\end{cases}.
\end{align}
The above equation indicates that Huber's minimax norm has a re-descending influence function and has a well defined anisotropic weighting function. This norm is widely used in image processing applications but has---to the best of our knowledge---not been used for face normal filtering yet \edit{and is therefore not included in Table~\ref{tab:sota}}.
       
\subsubsection{\edit{d) Yadav et al.~\cite{Yadav2017}}}
introduced a face normal filtering technique using a box filter as the anisotropic weighting function
\begin{align}
	\label{equ:yadav_envt_weight_fct}
	g_\sigma(x) = \begin{cases}
	1 & \edit{|x| < \sigma}\\
	0.1 & \text{otrw}.
	\end{cases}, && \text{with} &&  x = \angle(n_i, n_j),
\end{align}
where~$\angle(n_i, n_j)$ denotes the angle between the central normal~$n_i$ and it neighboring normal~$n_j$. The corresponding error norm and influence function can be derived as
\begin{equation}
	\label{equ:yadav_envt_rho}
	\rho_\sigma(x) = \int\limits_{0}^{x} x' g_\sigma(x')dx' = \begin{cases}
	x^2 & \edit{|x| < \sigma}\\
	\edit{0.1(x^2 + 9\sigma^2)}& \text{otrw}.
	\end{cases},
\end{equation}
\begin{equation}
	\label{equ:yadav_envt_psi}
	\psi_\sigma(x) = x g_\sigma(x) = \begin{cases}
	x & \edit{|x| < \sigma}\\
	0.1x& \text{otrw}.
	\end{cases}.
\end{equation}
From the above error norm and influence function, we can see that this filtering is using an error norm quite similar to the truncated quadratic error norm (see second row in Table~\ref{tab:estimators}) for the computation of the element-based normal voting tensor. The corresponding influence function is neither bounded nor re-descending, but the outlier effect will be quite minimal. This is because of the down-scaling of the argument in the influence function for bigger $x$. Therefore, the algorithm is able to preserve sharp features. However, it is less robust against high noise intensities because of the non-re-descending and unbounded influence function.

\subsubsection{\edit{e) Shen et al. \cite{Shen2004}}}
\edit{introduced the fuzzy vector median-based surface smoothing algorithm, which is quite similar to the algorithm of~\cite{Belyaev2001} (explained in paragraph~\textbf{a)} in the beginning of this section).} The anisotropic weighting function~$g_\sigma(x)$ is a Gaussian function as given in Equation~\rf{equ:belayev2002} and the input~$x$ is given as
\begin{equation*}
	x = \norm{n_j - n_{vd}},
\end{equation*}
where~$n_j$ represents neighboring normals to the processed central face~$f_i$ and the term~$n_{vd}$ performs \emph{vector directional median filtering} on the normal vectors including the central normal~$n_i$. Vector directional median filtering is an extension of \emph{median filtering} for multivariate data, see~\cite{Trahanias1993}, and can be computed as
\begin{equation}
	n_{vd} = \underset{n}{\operatorname{argmin}} \sum_{j\in \Omega_{vd}} \angle(n, n_j),
\end{equation}
where~$\angle(n, n_j)$ denotes the angle between~$n$ and~$n_j$ and the set~$\Omega_{vd}=\Omega_i\cup\{i\}$ consists of indices of the neighbor normals~$n_j$ together with the index~$i$ of the central normal~$n_i$.

The corresponding influence function will be re-descending as shown in Equations~\rf{equ:belayevRho} and~\rf{equ:belayevPsi}. The input argument of~$g_\sigma(x)$ is the Euclidean difference between the neighboring normals and their median. This method performs well in terms of feature preservation but is not robust during noise removal because of the unavailability of the spatial filter. As it is clear from Equations~\rf{equ:AnisotropicDiffusionEquation}, the anisotropic weighting function~$g_\sigma(x)$ is similar to the edge stopping function in the diffusion process. 

\subsubsection{\edit{f) Tasdizen et al.~\cite{Tasdizen2002}}}
apply---based on the relationship between bilateral filtering and non-linear diffusion~\cite{Barash2002}---the diffusion of face normals for filtering by using the Gaussian function as anisotropic weighting function. Curvature information is used as input~$x$ in this algorithm. Similar to the method of~\cite{Belyaev2001}, from Equations~\rf{equ:belayev2002},~\rf{equ:belayevRho}, and~\rf{equ:belayevPsi} it can be derived that this method also follows the Gaussian error norm and has a bounded, re-descending influence function, which helps preserving sharp features. However, due to unavailability of the spatial filter, this algorithm is not robust against significant noise.

\subsubsection{\edit{g) Centin et al.~\cite{Centin2018}}}
 also introduce a face normal diffusion method using the following anisotropic weighting function
\begin{align}
	g_\sigma(x) = \begin{cases}
		1 & \edit{|x| < \sigma}\\
		\frac{\sigma^2}{(\sigma - x)^2 + \sigma^2} & \text{otrw}.
	\end{cases}, && \text{ where} && x = \kappa \cdot \ell_{avg}.
\end{align}
The term $\kappa$ represents curvature information computed at each face by averaging the curvature at the corresponding vertices and~$\ell_{avg}$ represents the average edge length computed over the entire geometry. The corresponding influence function can be derived as
\begin{equation}
	\psi_\sigma(x) = xg_\sigma(x) = \begin{cases}
		x & \edit{|x| < \sigma}\\
		\frac{x\sigma^2}{(\sigma - x)^2 + \sigma^2} & \text{otrw}.
	\end{cases}.
\end{equation}
The above influence function is bounded and re-descending, which makes this algorithm effective in terms of feature-preservation. This method falls somewhere between the Lorentzian error norm (decaying of~$g_\sigma(x)$ for~${x\geq\sigma}$) and Huber's minimax error norm (constant~$g_\sigma(x)$ for~${x<\sigma}$).  Due to absence of a spatial filter, this algorithm is not robust against high intensities of noise.

\subsection{\edit{Bilateral normal filtering}}
\label{sec:bilMesh}
Bilateral normal filtering is one of the most effective and robust approaches for denoising of normals. In contrast to unilateral normal filtering, the weighting function in bilateral normal filtering consists of two different Gaussian kernels. As above, one kernel carries the anisotropic nature and is commonly known as range filter (we termed it anisotropic weighting function~$g_\sigma(x)$) while the other kernel is known as spatial kernel (given as~$f_{\sigma_d}(d)$ in Equation~\rf{equ:RhoSpatial}) and is isotropic in nature. 

\subsubsection{\edit{a) Zheng et al.~\cite{Zheng2011}}} \edit{define these kernels as:}
\begin{align}
\label{equ:meshBil}
	g_\sigma(x) = \exp\left( -\frac{x^2}{2\sigma^2}\right) && \text{and} && f_{\sigma_d}(d) = \exp\left( -\frac{d^2}{2\sigma_d^2}\right), 
\end{align} 
where~$\sigma_d$ is the average distance between neighboring faces and the central face. The input arguments~$x$ and~$d$ are defined as:
\begin{align*}
	x = \norm{n_i - n_j} && \text{and} && d = \norm{c_i-c_j},
\end{align*} 
where $c_i$ and $c_j$ are the centroids of the central face $f_i$ and the neighboring face~$f_j$ respectively. 

In the robust statistics framework, our main focus is the anisotropic weighting function~$g_\sigma(x)$, its corresponding error norm, and the corresponding influence function because $g_\sigma(x)$ is responsible for feature preservation. From Equations~\rf{equ:belayev2002},~\rf{equ:belayevRho}, and~\rf{equ:belayevPsi}, it is clear that the method of~\cite{Zheng2011} has a re-descending influence function (second last row of Table~\ref{tab:estimators}). Thereby, this algorithm is capable of preserving sharp features effectively and removes noise better compared to the algorithms mentioned above because of the utilized spatial filter~$f_{\sigma_d}(d)$.

\subsubsection{\edit{b) Zhang et al.~\cite{Zhang2015}}}
describes a procedure of guided mesh normal filtering following the Gaussian error norm and uses the same spatial filter as the method of~\cite{Zheng2011} presented above. The guided mesh normal is based on a joint bilateral filter, where an anisotropic weighting function (range kernel) works on the guidance signal. That is, the input variable~$x$ is defined as:
\begin{equation}
\label{equ:guidedMesh}
x = \norm{G_i - G_j},
\end{equation}
where~$G_i$ and~$G_j$ are the guidance normals, which are computed by averaging similar normals in the respective neighborhood.

\subsubsection{\edit{c) Yadav et al.~\cite{Yadav2018}}} introduce a bilateral normal filtering using the following anisotropic \edit{weighting} function:
\begin{align}
	g_\sigma(x) = \begin{cases}
	\frac{1}{2} \left[1-\left(\frac{x}{\sigma} \right)\edit{^2} \right]^2 & \edit{|x| \leq \sigma}\\
	0 & \text{otrw}.
	\end{cases}, && \text{where} && x = \norm{n_i-n_j}.
\end{align}
The above function is known as Tukey's bi-weight function~\cite{Beaton1974}. The spatial filter~$f_{\sigma_d}(d)$ is a Gaussian function similar to that used in the method of~\cite{Zheng2011} as described above.
In the robust statistics framework, the corresponding influence function and error norm can be derived as
\begin{equation}
	\psi_\sigma(x) = x g_\sigma(x) = \begin{cases}
		\frac{x}{2} \left[1-\left(\frac{x}{\sigma} \right)\edit{^2} \right]^2 & \edit{|x| < \sigma}\\
		0 & \text{otrw}.
 	\end{cases},
\end{equation} 	
\begin{equation}
	\label{equ:TukeyErrorNorm}
	\rho_\sigma(x) = \int\limits_{0}^{x} x' g_\sigma(x') dx' = \begin{cases}
		\frac{x^2}{\sigma^2} - \frac{x^4}{\sigma^4} + \frac{x^6}{3\sigma^6}  & \edit{|x|} < \sigma\\
		\frac{1}{3} & \text{otrw}.
	\end{cases}.
\end{equation}
From the influence function and error norm, it is clear that Tukey's bi-weight function is more robust compared to the Gaussian function in terms of feature preservation because it strictly cuts off outliers with respect to the user-chosen parameter~$\sigma$. Also, the Gaussian spatial filter helps to remove noise components effectively.

\section{Point Set Surface Denoising in the Robust Statistics Framework}
\label{sec:PointSetSurfaceDenoising}

In this section, we will shift our focus slightly. Instead of an input mesh $\mathcal{M}$, we will now consider a point set sample of a surface (PSS) as input. Thus, we are only given vertices ${P=\{p_i\}_{i\in I}\subseteq\mathbb{R}^3}$ with corresponding normals $N=\{n_i\}_{i\in I}$, i.e.\@ compared to the above we cannot use edges to induce connectivity between the vertices nor can we use the area of faces as weighting terms in the filtering process.

Despite these challenges, a multitude of procedures and algorithms have been proposed for the denoising of PSS. This is mostly due to two advantages of PSS over meshes. First, point sets are often the raw output of 3D acquisition devices and processes. Thus, if an algorithm is available to work on a PSS, it can be directly---possibly even on site---applied to the acquired data. Second, as there is no connectivity information in the point set, no such data has to be stored, which amounts to significantly lower storage costs compared to meshes. Furthermore, no topological problems---like non-manifold edges or fold-overs---and no numerical problems---like slivers---are introduced as the PSS only gives an implicit handle on the underlying surface geometry.

In the following, we will focus on adaptations of face normal filtering algorithms from meshes to point sets as well as on original methods proposed directly in the PSS setting. Note that any method on point sets can easily be applied to the meshed setting by simply disregarding the edge and face connectivity information.

\subsection{\edit{Unilateral normal filtering}}

As for meshes, we will first focus on unilateral normal filtering procedures. These do not use a specific spatial filter, i.e.~$f_{\sigma_d}(d)\equiv1$. This makes them less robust against moderate or high levels of noise.

\subsubsection{\edit{a) Öztireli et al.\cite{2009Oztireli_FeatPresPCNonLinKernRegr}}}
introduced a modification of the Moving Least Squares (MLS) procedure~\cite{alexa2003computing} aiming at the integration of feature-preservation into the MLS pipeline. Their core objective is an iterative minimization and can be understood as iterative trilateral filtering, as it makes use of three types of weights. The first one is isotropic in nature and appears as~$ \mathcal{C}^3 $ continuous polynomial approximation of the Gaussian, i.e.\@ 
\begin{equation}
f_i(p) = \left( 1 - \frac{\norm{p - p_i}}{h_i^2}^2 \right)^4
\end{equation}
where the argument~$ p $ is some point (not necessarily from~$P$), as the objective is an implicit, signed distance function.  The value~$ h_i $ is a weight adapting the local density, chosen within a range from~$ 1.4 $ to~$ 4 $ as experimentally evaluated by the authors~\cite{2009Oztireli_FeatPresPCNonLinKernRegr}. For the second weighting term---using the height over an estimated hyperplane at~$p$ and thus capturing both isotropic and anisotropic quantities---the authors discuss M-estimators and include the Gaussian error norm and its respective Gaussian error weight, see Equation~\rf{equ:belayev2002}, into their optimization problem. The arguments are
\begin{align*}
d = y_i - \tilde{\eta}^{k-1}(p_i) && \text{and} && \sigma_d = \frac{h_i}{2},
\end{align*}
with~$ y_i $ the heights of the samples~$ p_i $ taken over the local least-squared best fitting hyperplane, and~$ \tilde{\eta}^{k-1} $ the corresponding local approximation. The value for~$ \sigma_d $ is set fix throughout the whole paper by the authors. The third and final weighting terms are anisotropic and make use of a Gaussian function with arguments
\begin{align*}
x = \norm{\nabla \eta^k(p) - n_i} && \text{and} && \sigma \in \mathbb{R},
\end{align*}
where~$ \eta $ is an implicit, signed distance function as main objective,~$ p $ some point at which we want to evaluate the function~$ \eta $,~$ n_i $ the normal at sample point~$ p_i $, and~$ \sigma $ a parameter that regulates the sharpness where typical choices range from~$ 0.5 $ up to~$ 1.5 $. This last weighting term penalizes the deviation of normals when we reach sharp features. The influence function and error norm are of Gaussian nature and are derived in Equations~\rf{equ:belayevPsi} and~\rf{equ:belayevRho}.
The assembled combination yields a robust implicit surface definition via MLS, which can represent both smooth surface patches and sharp features and was coined robust implicit MLS (RIMLS). Similar to Method~\cite{Belyaev2001}, this algorithm is capable of retaining and enhancing sharp features. However, the unavailability of a spatial filter~$f_{\sigma_d}(d)$ makes the algorithm less effective against moderate and high levels of noise.

\subsubsection{\edit{b) Mattei and Castrodad~\cite{2016Mattei_PCDenoisingMRPCA}}}
start their paper with the assertion that the Principal Component Analysis~(PCA) operation for the estimation of local reference planes is not robust. They proceed to construct a moving robust PCA~(MRPCA). Their main ingredient of interest in the given context is a minimization problem, which makes use of anisotropic weights determined via the Gaussian weight function as given in Equation~\rf{equ:belayev2002} with arguments
\begin{align*}
x = \arccos(n_i \cdot n_j) && \text{and} && \sigma \in \mathbb{R},
\end{align*}
where~$ n_i, n_j $ are the unit normals at the considered point~$ p_i $ and at one of its neighbors~$ p_j $ (with a $k$-nearest neighborhood utilized). Furthermore, $ \sigma $ is a bandwidth parameter affecting the reconstruction of sharp features. The authors propose values of~$ {\sigma \in (\pi / 12, \pi / 6)} $. Using this anisotropic weight function yields the Gaussian error norm along with its re-descending influence function as given in Equations~\rf{equ:belayevPsi} and~\rf{equ:belayevRho}. Similar to Method~\cite{Belyaev2001}, this algorithm is capable of retaining and enhancing sharp features. However, the unavailability of a spatial filter~$f_{\sigma_d}(d)$ makes the algorithm less effective against moderate and high levels of noise.   

\subsection{\edit{Bilateral normal filtering}}

We will now turn to bilateral normal filtering procedures for PSS. These use two different weighting kernels. As for meshes, one kernel carries the anisotropic nature while the other one of isotropic behavior.

\subsubsection{\edit{a) Li et al.~\cite{2009Li_FADenoisePSS}}}
presented one of the first approaches applying bilateral filtering to PSS. The authors first estimate the likelihood $\ell_i$ that a given sample point~${p_i\in P}$ is close to the underlying surface geometry. They propose to compute $\ell_i$ based on the MLS technique of~\cite{alexa2003computing}.
The normal denoising utilizes the bilateral filtering scheme, which includes a Gaussian weighting (following Equation~\rf{equ:belayev2002}) as a spatial filter $f_{\sigma_d}(d)$ with the following input arguments in the isotropic setting
\begin{align*}
	d = \norm{p_i - p_j} && \text{and} && \sigma_d = \frac{r}{{2}},
\end{align*}
and another Gaussian weighting function $g_\sigma(x)$ in the anisotropic setting with following input arguments 
\begin{align*}
	x = \arccos(n_i \cdot n_j) && \text{and} && \sigma \in \mathbb{R},
\end{align*}
the latter chosen to be the standard deviation of the normal variation given in~$x$. Here,~$ r $ is the radius of the enclosing sphere of the geometric neighborhood~$\Omega_i$. Observe that the values presented here differ from those given in~\cite{2009Li_FADenoisePSS}, because we adjust them to fit the Gaussian given in Equation~\rf{equ:belayev2002}. Lastly, the closeness of the point~$p_i$ to the underlying surface, measured by~$\ell_i$, the feature intensity, and the bilateral filtering for normals are used in a final sample point filtering step to remove noise from the PSS. The mentioned method follows the Gaussian error norm similar to the bilateral normal filtering of \cite{Zheng2011}. As shown in Equation~\rf{equ:belayevPsi}, the applied anisotropic weighting function $g_\sigma(x)$ has a re-descending and bounded influence function, which makes the algorithm robust in terms of feature preservation and also the availability of the spatial filter~$f_{\sigma_d}(d)$ ensures the effectiveness towards different levels of noise.  

\subsubsection{\edit{b) Zheng et al.~\cite{2017Zheng_GuidedPCDenoising}}}
proposed a four-stage method for point set denoising. It consists of sharp feature detection, multiple normals computation, guided normal filtering, and point updating. Concerning the feature detection, the authors provide a two-step procedure: feature candidate detection and feature point selection. The former is to find the global feature structure and utilizes the framework of robust statistics. Namely, after a first computation of normals using PCA, the normal similarity is evaluated via the Gaussian weight function, see Equation~\rf{equ:belayev2002}, with arguments
\begin{align*}
	x = \norm{n_i - n_j} && \text{and} && \sigma \in \mathbb{R},
\end{align*}
with a user-given angle-threshold~$\sigma$, which ranges from~$0.05$ to~$0.3$ in the experiments of the authors,~$ n_i $ the normal at the considered point and~$ n_j $ the normal at one of its neighbors, while using the $k$-nearest neighbors as neighborhood notion. In contrast to the single normal used in the normal similarity described above, the authors of~\cite{2017Zheng_GuidedPCDenoising} attach bundles---a multitude of normals---to every point. A comparable approach is then chosen to estimate averaged normals utilizing spatial weights evaluated once more via the Gaussian weight function~\rf{equ:belayev2002} with arguments
\begin{align*}
	d = \norm{p_i-p_j} && \text{and} && \sigma_d \in \mathbb{R},
\end{align*}
with~$\sigma_d$ ranging from~$ 0.1 $ to~$ 0.5 $ in the \edit{authors'} experiments. Finally, both weightings are combined in the actual bilateral normal filtering. This method is an extension of guided mesh normal filtering~\cite{Zhang2015}, which we have mentioned in Equation~\rf{equ:guidedMesh}. From the explanation for guided mesh normal filtering in Section~\ref{sec:bilMesh}, it is clear that this method also follows the Gaussian error norm along with a bounded and re-descending influence function and has similar robustness in terms of feature preservation and noise-removal. The computation of guided normals makes this algorithm slightly better compared to bilateral normal filtering.

\subsubsection{\edit{c) Park et al.~\cite{2013Park_FAFilteringPSS}}}
proposed a three-staged point set filtering approach including feature detection, normal re-calculation, and a point position update. Their feature detection tensor, adaptive sub-neighborhood, and point update all use the Gaussian weighting function given in Equation~\rf{equ:belayev2002}, where for the first two, the arguments are of anisotropic nature given as
\begin{align*}
	x = \sqrt{s^2 + c \kappa^2} && \text{and} && \sigma \in \mathbb{R},
\end{align*}
with a prescribed constant~$ c $,~$ \sigma $ set by the authors to the neighborhood range, which is~$ 4\delta $ with~$ \delta $ the arithmetic mean of all distances of the points to their closest neighbors respectively. The value~$ s $ represents the arc-length on the tangent plane and~$ \kappa $ the curvature obtained by the circle, which goes through both the center point~$ p_i $ and its considered neighbor~$ p_j $ and which is also tangent to the attached normals~$ n_i $ and~$ n_j $. These normals are calculated via an initial normal estimation following~\cite{1992Hoppe_SurfReconUnorgPts}.
To compute the feature detection tensor, the method uses a Gaussian function as the anisotropic weighting, which has a re-descending influence function~$ \psi_\sigma $ and a derived Gaussian error norm~$ \rho_\sigma $ as given in Equations~\rf{equ:belayevPsi} and~\rf{equ:belayevRho} respectively. In terms of feature sensitivity, it will be as effective as MRPCA. However, this algorithm is not robust against moderate and high levels of noise.

\subsubsection{\edit{d) Digne and de Francis~\cite{2017Digne_BilateralFilterPC}}}
\edit{proposed an extension of the bilateral filtering on meshes to points via a parallel implementation of~\cite{2003Fleishman_BilMeshDenoising} using points.} The whole procedure consists of a point update using non-oriented normals and utilizes Gaussian weights, Equation~\rf{equ:belayev2002}, twice, with isotropic
\begin{align*}
	d = \norm{p_i-p_j} && \text{and} && \sigma_d = \frac{1}{3} r,
\end{align*}
 and anisotropic arguments
\begin{align*}
	x = \left|n_i \cdot (p_j - p_i)\right| && \text{and} && \sigma = \frac{1}{3} r' ,
\end{align*}
with user given radii~$ r $ and~$ r' $. If these are not given, the authors use a heuristic and set~$ {r = \ell \sqrt{20 / |P|}} $, where~$ \ell $ denotes the size of the bounding box and~$ |P| $ the number of vertices. The values~$ \sigma_d $ and~$ \sigma $ are set to be equal in this case. The point~$ p_i $ is the one considered to be updated and~$ p_j $ represents one of its neighbors within a geometrical neighborhood~$ \Omega_i $. The weights determined by~$ f_{\sigma_d} $ measure the spatial distance and those by~$ g_\sigma $ evaluate the distance of neighbors to the plane spanned by the point~$ p_i $ and its normal. As the weights are of Gaussian nature, we can derive the influence function and Gaussian error norm given in Equations~\rf{equ:belayevPsi} and~\rf{equ:belayevRho}. In terms of feature-preservation and noise-removal, this algorithm will be as effective as bilateral normal filtering~\cite{Zheng2011} as both of them are using same robust error norm with a slightly different input argument.  

\subsubsection{\edit{e) Zheng et al.~\cite{2018Zhen_RollingNormalFilterPC}}}
propose an iterative two-staged denoising algorithm which---in contrast to most methods---smooths out smaller features while preserving larger ones. The iterative normal filtering (with initial normals obtained via~\cite{1992Hoppe_SurfReconUnorgPts}) and the following point position update (solved iteratively via gradient descent) make use of the Gaussian weighting, Equation~\rf{equ:belayev2002}, with the isotropic arguments
\begin{align*}
	d = \norm{p_i - p_j} && \text{and} && \sigma_d \in \mathbb{R}
\end{align*}
and the anisotropic arguments
\begin{align*}
	x = \norm{n_i - n_j} && \text{and} && \sigma \in \mathbb{R},
\end{align*}
where~$ \sigma_d \in [0.01, 0.5] $ and~$ \sigma \in [0.1, 0.5] $ given in the authors' experiments,~$ p_i $ the considered point,~$ p_j $ representing its neighbor ($ k $-nearest neighbors are used), and~$ {n_i, n_j} $ the respective normals. Consequently, the evaluation is similar and on the one hand uses spatial distances of points while on the other hand using closeness of normals. The used Gaussian weights yield the influence function and Gaussian error norm given in Equations~\rf{equ:belayevPsi} and~\rf{equ:belayevRho}, which make this algorithm robust in terms of feature preservation and noise-removal. One of the key benefits of this algorithm is that by adjusting the parameter~$\sigma$, different levels of features can be smoothed out effectively. \edit{An even more robust version, utilizing the same weighting terms as given above, is discussed in~\cite{sun2019reliable}.}

\subsubsection{\edit{f) Yadav et al.~\cite{Yadav2018_VNVT}}}
offers an extension of~\cite{Yadav2017} to point sets. The proposed iterative scheme consists of the following three stages: normal filtering, feature detection, and vertex update. The first two make use of a similar box filter as given in Equation~\rf{equ:yadav_envt_weight_fct}, here given as
\begin{align*}
	g_\sigma(x) = \begin{cases}
	1 & x \leq \sigma \\
	0 & \text{otrw}.
	\end{cases}
\end{align*}
with input arguments
\begin{align*}
	x = \arccos(n_i \cdot n_j) && \text{and} && \sigma \in \mathbb{R},
\end{align*}
\edit{where~$ {n_i, n_j} $ are unit-length normals and~$ \sigma $ is an angle-threshold for the neighbor selection (chosen by the user).} The deviation from the weighting defined in~\cite{Yadav2017} is because vertex normals are more sensitive to noise compared to face normals. Similar to the influence function and error norm derived in Equations~\rf{equ:yadav_envt_psi} and~\rf{equ:yadav_envt_rho}, the anisotropic weights given above yield an influence function of
\begin{equation*}
\label{equ:yadav_vnvt_psi}
\psi_\sigma(x) = x g_\sigma(x) = \begin{cases}
x & \edit{|x| < \sigma}\\
0 & \text{otrw}.
\end{cases}
\end{equation*}
and an error norm of
\begin{equation*}
\label{equ:yadav_vnvt_rho}
\rho_\sigma(x) = \int\limits_{0}^{x} x' g_\sigma(x')dx' = \begin{cases}
x^2 & \edit{|x| < \sigma}\\
0 & \text{otrw}.
\end{cases}.
\end{equation*}
The latter is a version of the truncated quadratic error norm, see the second row of Table~\ref{tab:estimators}. In contrast to~\cite{Yadav2017}, the influence function is both bounded and re-descending~(${ \psi \to 0 }$ when~${ x \to \infty }$). The impact of outliers is therefore kept small as it scales down for larger arguments~$ x $ and feature preservation is yielded. However, the performance of this algorithm is not optimal in the presence of moderate and high levels of noise due to the unavailability of a spatial filter~$f_{\sigma_d}(d)$.    

\subsubsection{\edit{Discussion: Local vs. Global Weighting}}
\edit{Note that out of the methods for point set surface denoising presented here, only~\cite{2009Oztireli_FeatPresPCNonLinKernRegr} utilizes a local vertex-based weight~$\sigma_d$. In contrast, methods~\cite{2009Li_FADenoisePSS,2017Zheng_GuidedPCDenoising,2017Digne_BilateralFilterPC,2018Zhen_RollingNormalFilterPC} use global weighting terms~$\sigma_d$. While localized terms can capture features on a finer level, they are harder to calibrate than global parameters. Furthermore, an implicit assumption of many algorithms is a noisy but uniformly dense sampling as input. Handling non-uniform densities requires additional work, see~\cite{skrodzki2018directional}. Finally, if the features of the input geometry are of comparable size, a global parameter is sufficient to capture them while still removing noise. Hence, most algorithms reduce to simple global parameters.}

\section{Experiments and Results}
\begin{figure}

	\centering
	\subfloat[Original]{\includegraphics[width=0.4\textwidth]{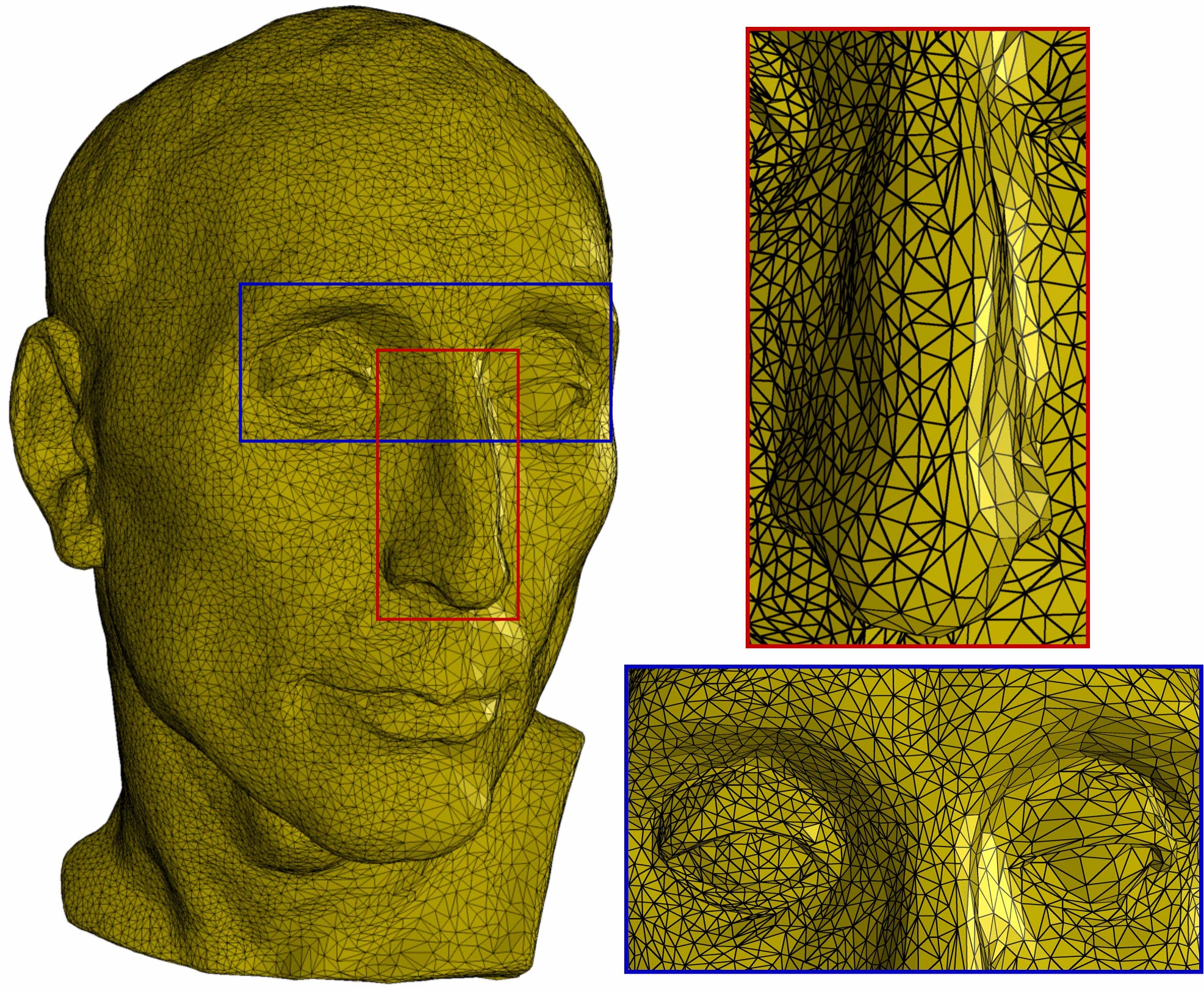}} \hfill
		 \subfloat[Noisy]{\includegraphics[width=0.4\textwidth]{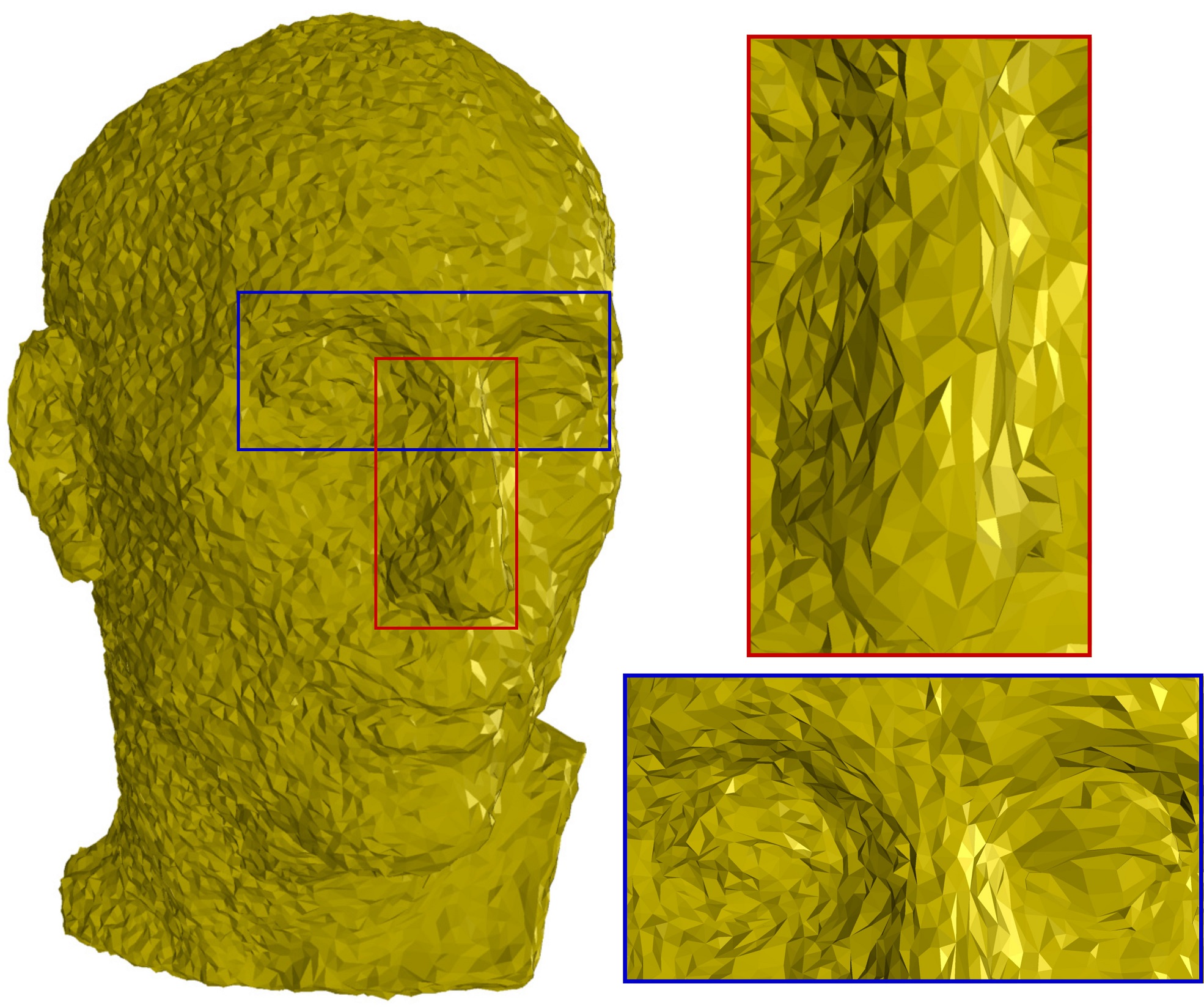}} \\
		\subfloat[$L_2$-norm \cite{Yagou2002}]{\includegraphics[width=0.4\textwidth]{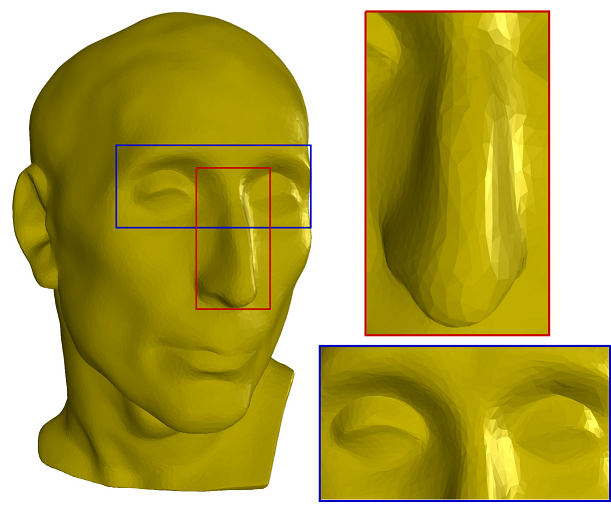}} \hfill
		 \subfloat[Truncated $L_2$-norm \cite{Yadav2017}]{\includegraphics[width=0.4\textwidth]{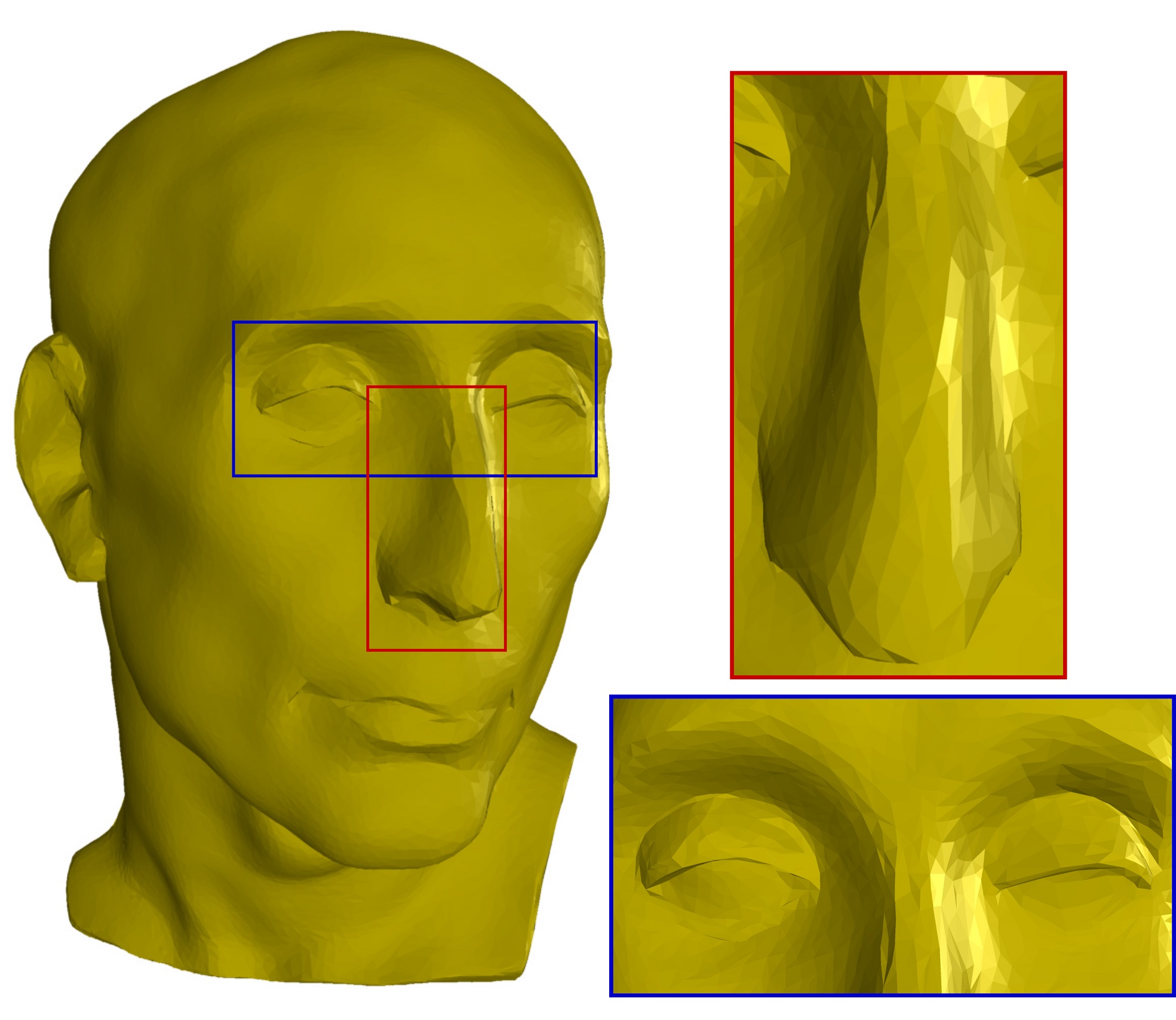}}\\
		\subfloat[Gaussian-norm \cite{Belyaev2001}]{\includegraphics[width=0.4\textwidth]{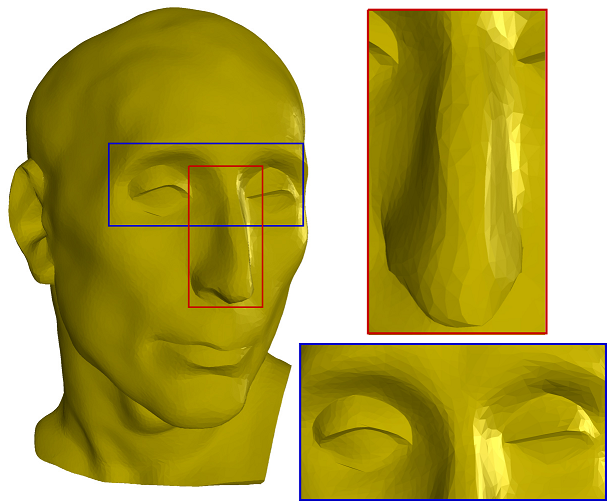}}\hfill
		\subfloat[Gaussian-norm with spatial filter \cite{Zheng2011}]{\includegraphics[width=0.4\textwidth]{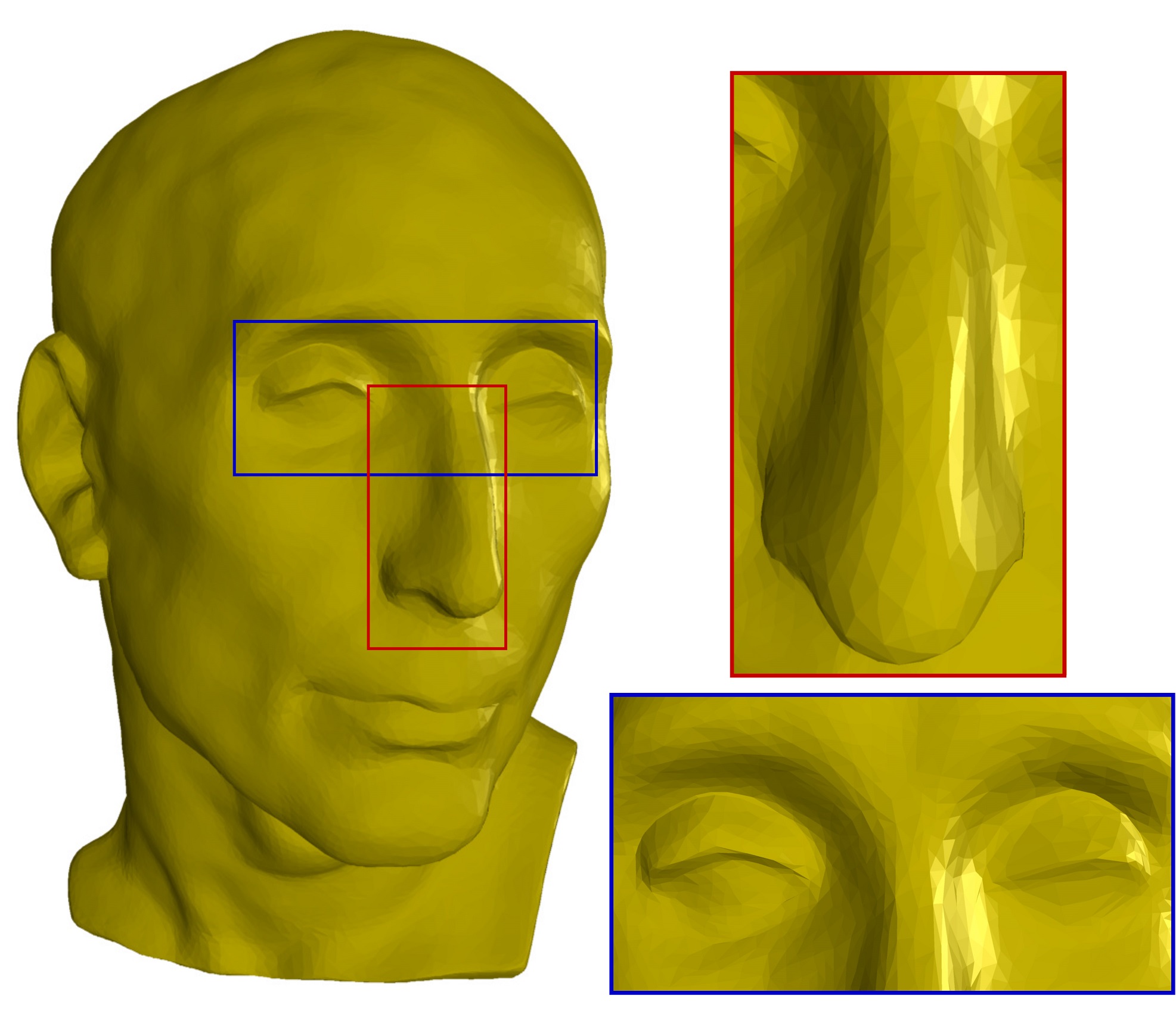}} \\
		\subfloat[Huber's minimax \cite{Centin2018}]{\includegraphics[width=0.4\textwidth]{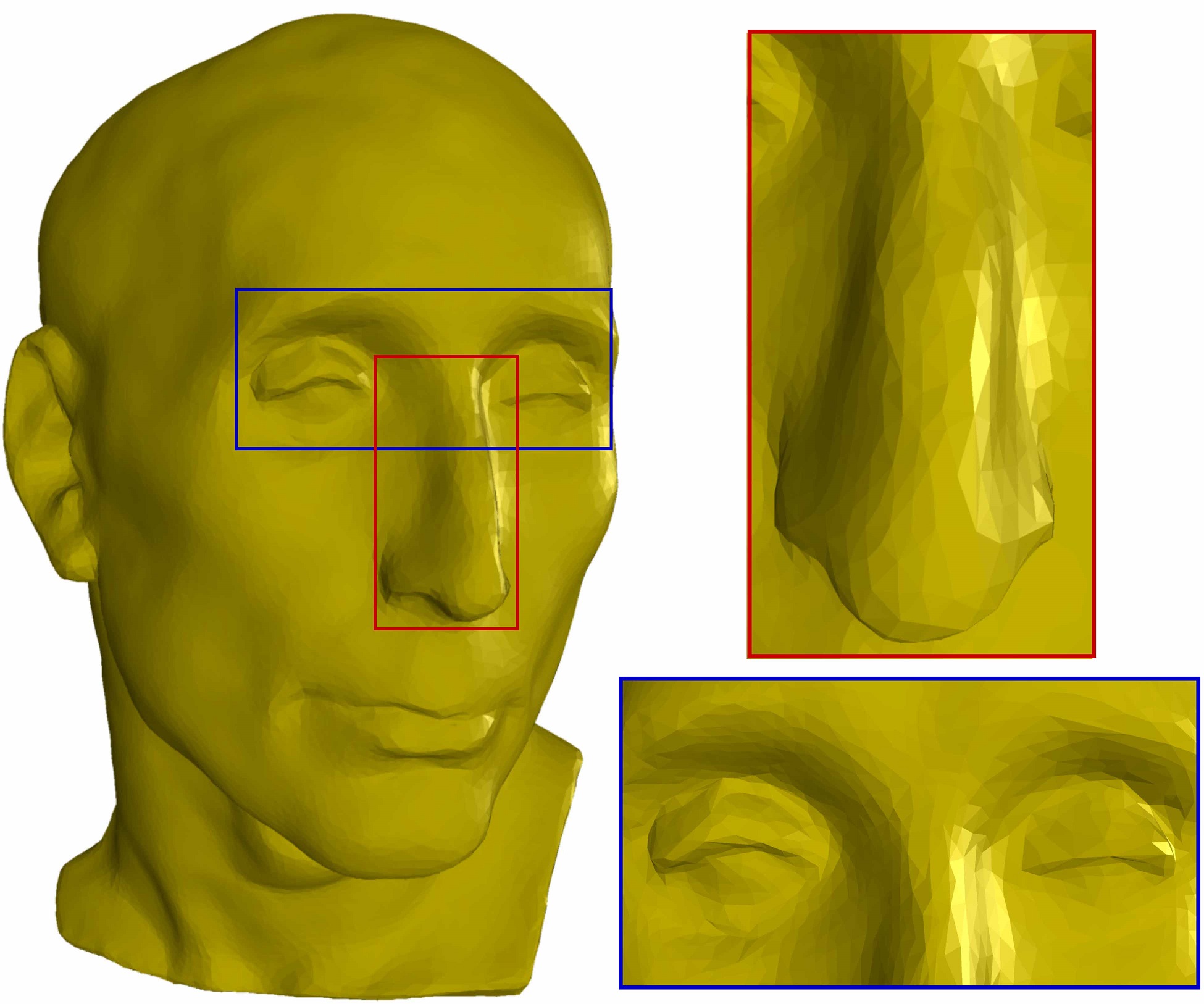}} \hfill
		\subfloat[Tukey's-norm \cite{Yadav2018}]{\includegraphics[width=0.4\textwidth]{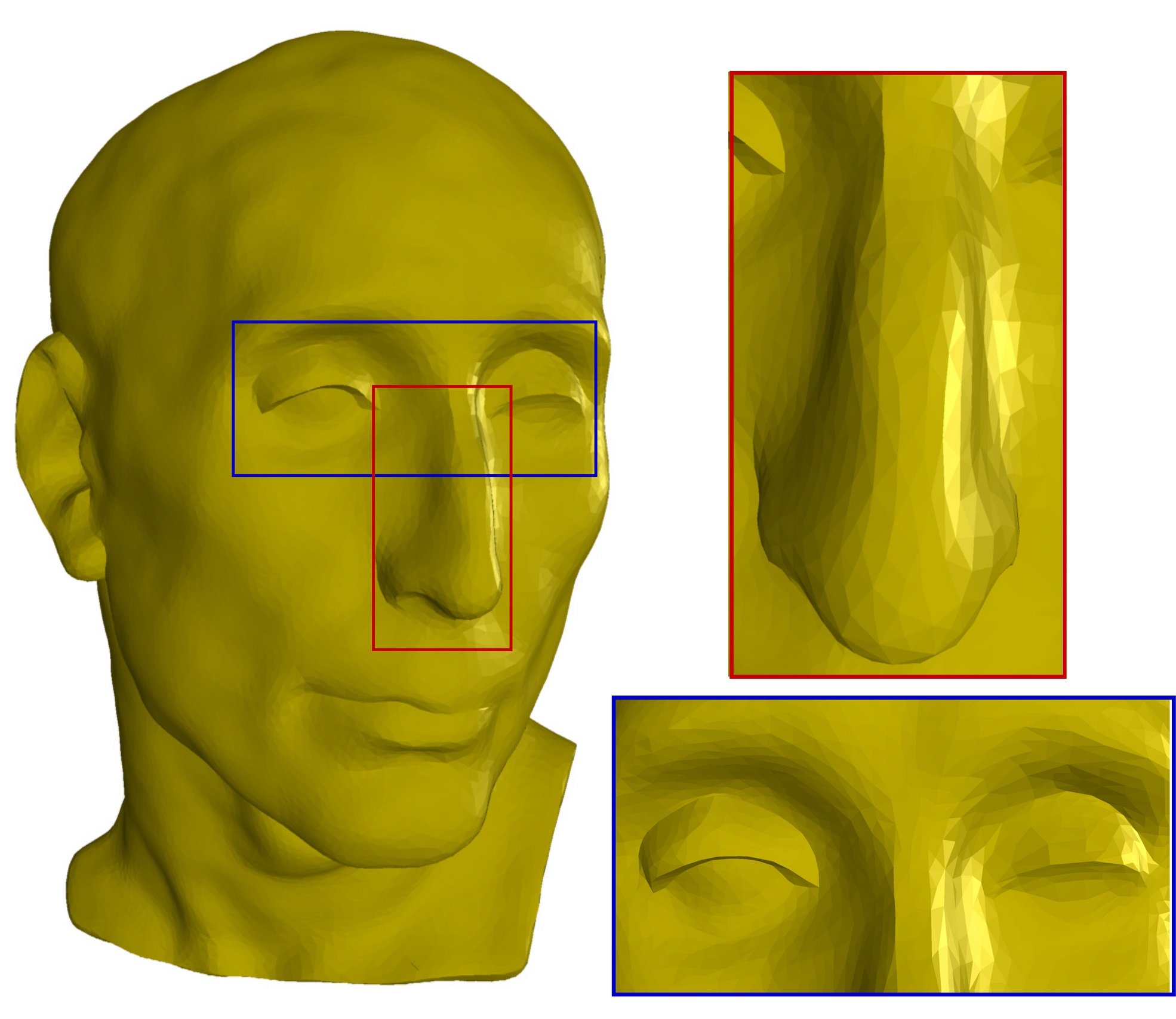}} 

\caption{Nicola model corrupted with a Gaussian noise ($\sigma_n = 0.2l_e$) in random direction. Images (c) to (h) show the results produced by state-of-the-art methods, which are using different robust error norms (see Table~\ref{tab:estimators}).}
\label{fig:nicola}
\end{figure}

In this section, we present experimental results regarding the state-of-the-art methods as listed in the previous sections, which are using different robust error norms. We have chosen two different models (CAD and CAGD) with different levels of noise. Figure~\ref{fig:nicola} shows the Nicola model corrupted with a moderate level of Gaussian noise (standard deviation~${\sigma_n = 0.2\ell_e}$, where~$\ell_e$ is the average edge length). Using this model, we show the capability of feature-preservation with the usage of different error norms. As shown in Figure~\ref{fig:nicola}, the~$L_2$-norm is not effective in terms of feature preservation (blurred eye region) because of the linear influence function and also as it is not bounded. The truncated~$L_2$-norm preserves features in the eye region better compared to the~$L_2$-norm as it has a truncated linear influence function. Figures~\ref{fig:nicola}(e) and~\ref{fig:nicola}(f) show the outputs of using the Gaussian norm without and with spatial filter, respectively. The Gaussian error norm has a re-descending influence function, which makes the algorithm more effective compared to the~$L_2$ related norms. The spatial filter is helping to remove noise effectively (eye and nose regions). Huber's minimax (Figure~\ref{fig:nicola}(e)) and the Gaussian error norm (Figure~\ref{fig:nicola}(g)) have quite similar outputs as they have re-descending influence functions and do not use spatial filters. Figure~\ref{fig:nicola}(h) shows the output of using Tukey's error norm, which has a sharper cut-off in the influence function compared to the Gaussian error norm. Therefore, feature-preservation is better compared to other norms mentioned and the spatial filter is helping to remove noise components effectively.

\begin{figure}
	\centering
	\subfloat[Original]{\includegraphics[width=0.4\textwidth]{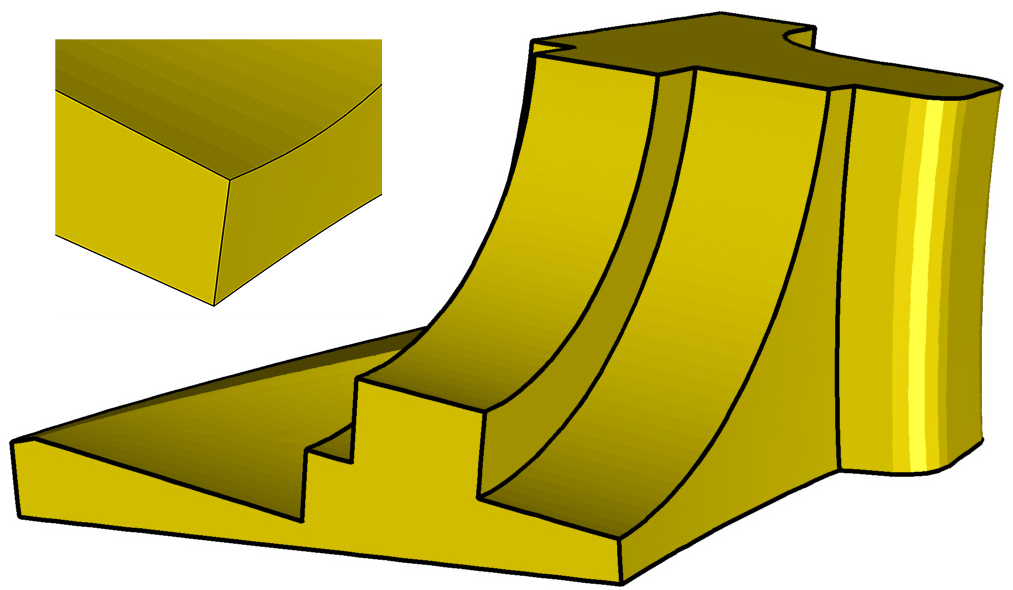}}
	\hfill
	\subfloat[Noisy]{\includegraphics[width=0.4\textwidth]{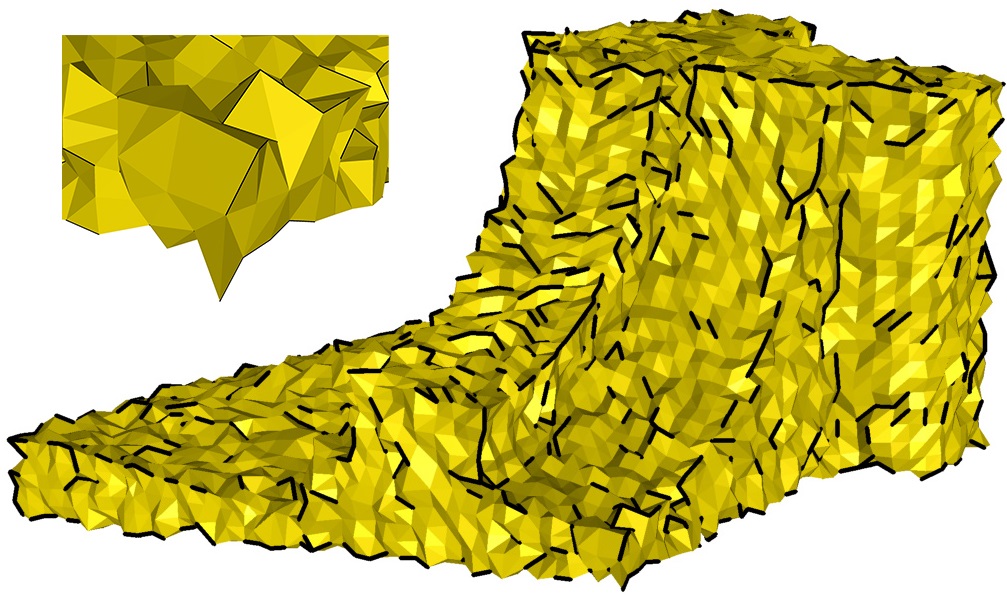}}\\
	\subfloat[$L_2$-norm \cite{Yagou2002}]{\includegraphics[width=0.4\textwidth]{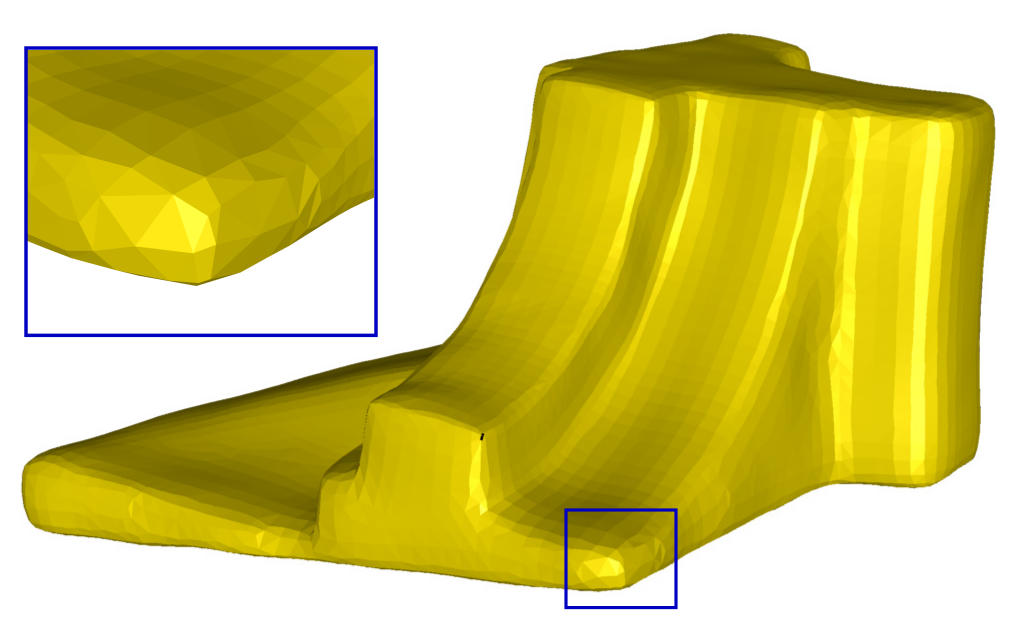}} 
	\hfill	
	\subfloat[Truncated $L_2$-norm \cite{Yadav2017}]{\includegraphics[width=0.4\textwidth]{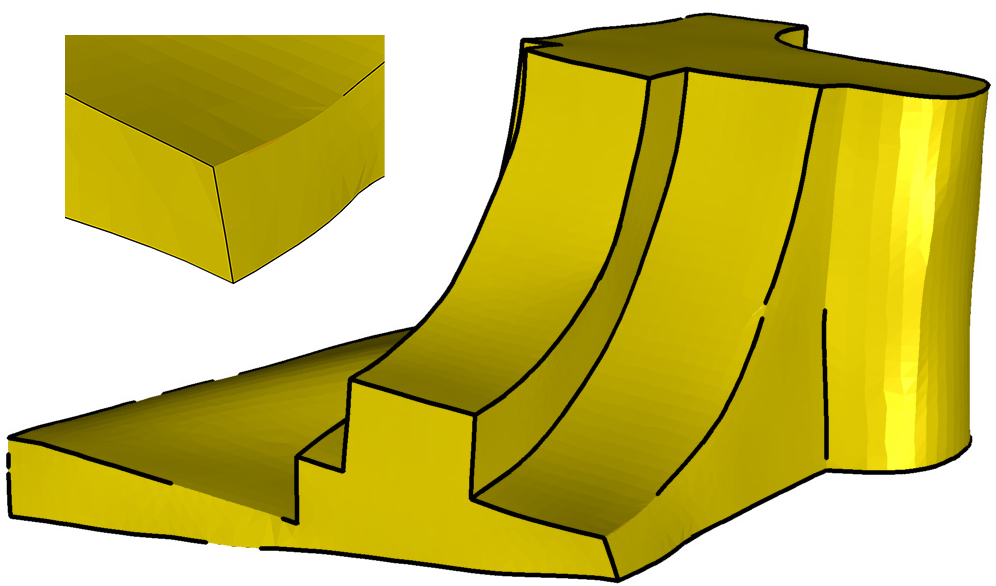}}\\
	\subfloat[Gaussian-norm \cite{Belyaev2001}]{\includegraphics[width=0.4\textwidth]{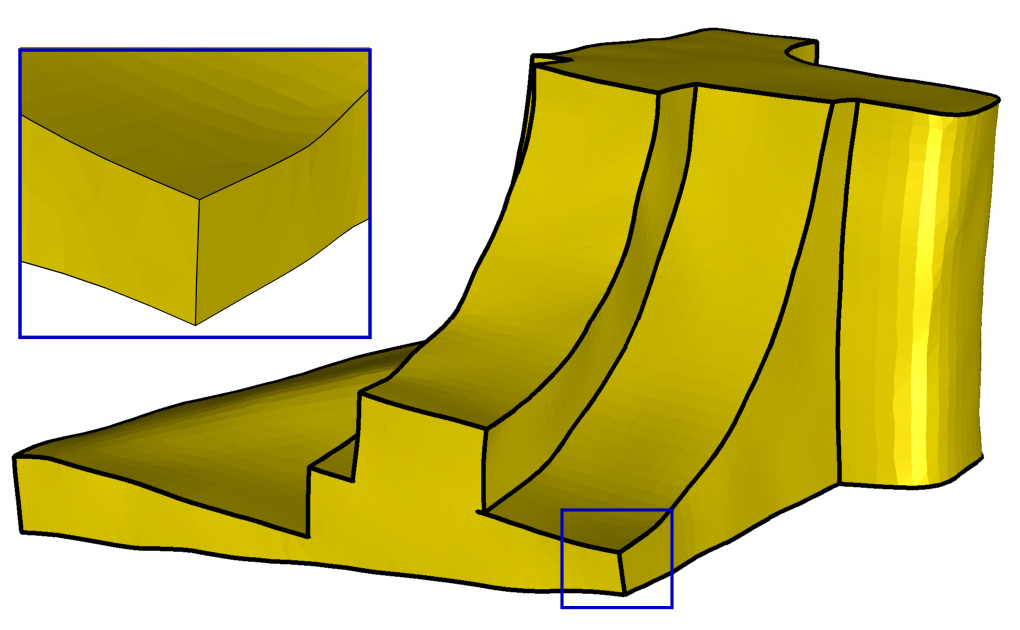}}
	\hfill
	\subfloat[Gaussian-norm with spatial filter \cite{Zheng2011}]{\includegraphics[width=0.4\textwidth]{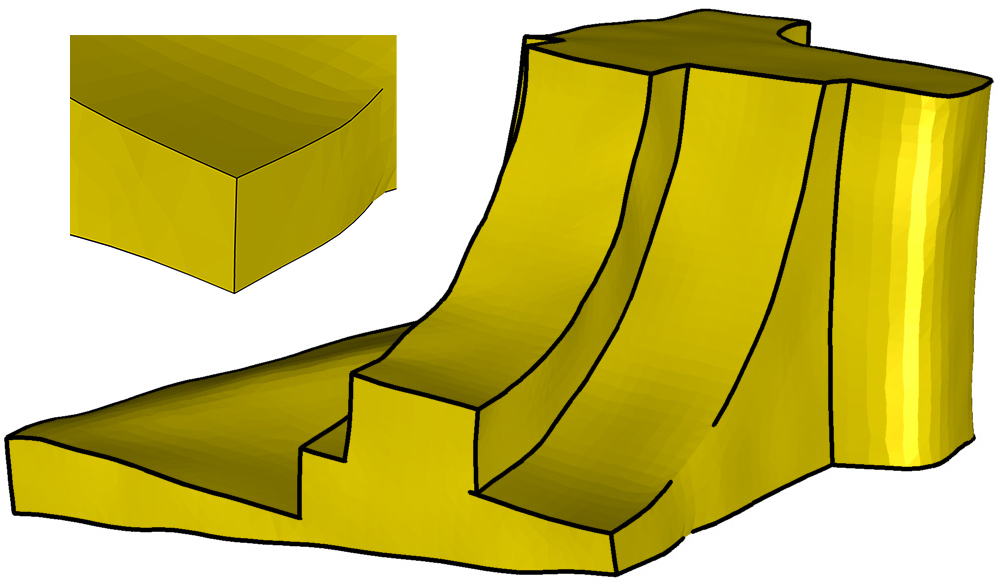}}\\
	\subfloat[Huber's minimax \cite{Centin2018}]{\includegraphics[width=0.4\textwidth]{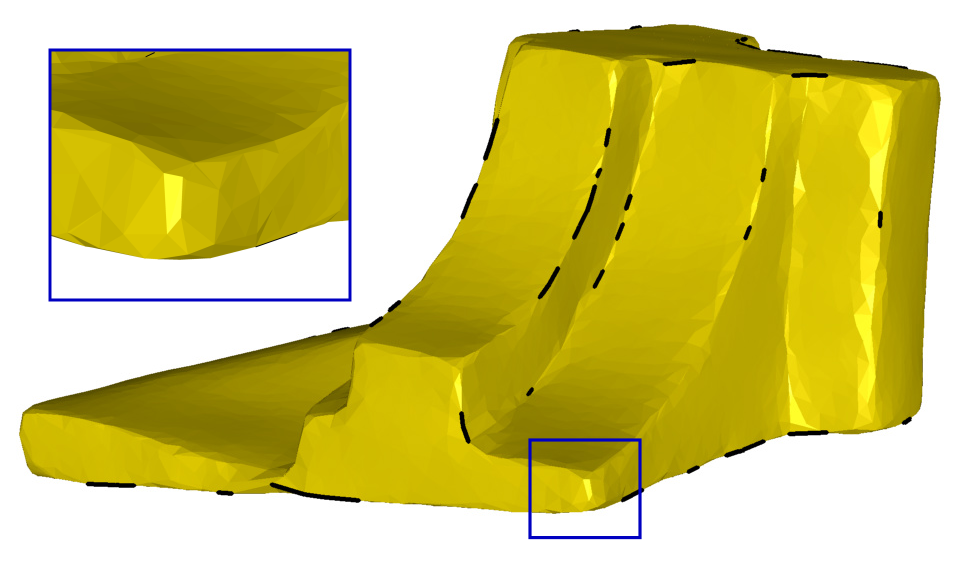}}
	\hfill
	\subfloat[Tukey's-norm \cite{Yadav2018}]{\includegraphics[width=0.4\textwidth]{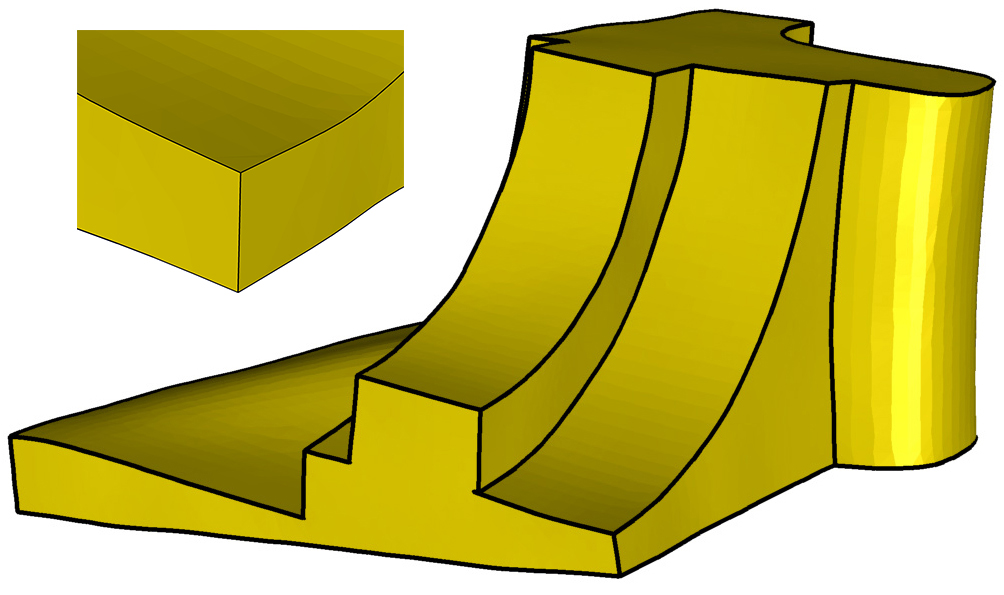}}
	\caption{The Fandisk model corrupted with a Gaussian noise~($\sigma_n = 0.3\ell_e$) in random direction. Figures (c) to (h) show the results produced by state-of-the-art methods, which are using different robust error norms (see Table~\ref{tab:estimators}). The black curve highlights sharp edge information in the geometries and is detected using a dihedral angle threshold of~${\theta = 70^\circ}$.}
	\label{fig:fandisk}%
\end{figure}

Figure~\ref{fig:fandisk} shows the robustness of the mentioned norm against high level of noise. The Fandisk model is corrupted  with a Gaussian noise~($\sigma_n = 0.3\ell_e$) in random direction. As it is shown, $L_2$ and Huber's minimax norms are able to remove the noise components effectively but feature-preservation is not effective. In case of the Gaussian error norm, the spatial filter removes different components of noise including low frequency ripples. However, the truncated $L_2$-norm is able to remove low frequency components by introducing an additional processing step (binary optimization) in the pipeline. The algorithm \cite{Yadav2018} uses Tukey's error norm, which helps to preserve features effectively and the spatial filter removes the noise components.   

\begin{sidewaystable} 
\begin{tabular}{lcc|ccc|cc}
	\multicolumn{2}{c}{Method} & Section & Input & Error Norm  & Spatial Weights & Feature-preservation & Noise-removal \\
	\hline
	Belyaev and Ohtake, 2001 & \cite{Belyaev2001} & 3.1 a & mesh & Gaussian & No& good & ok\\
	Yogou et al., 2002 & \cite{Yagou2002} & 3.1 b & mesh & $L_1$ and $L_2$ & No& ok & ok\\
	Yadav et al., 2017 & \cite{Yadav2017}& 3.1 d & mesh & Truncated $L_2$ & No&good & ok\\
	Shen and Barner, 2004 & \cite{Shen2004} & 3.1 e & mesh & Gaussian & No&good & ok\\
	Tasdizen et al., 2002 & \cite{Tasdizen2002} & 3.1 f & mesh & Gaussian  & No &good & ok\\
	Centin and Signoroni, 2018 & \cite{Centin2018}& 3.1 g & mesh & Huber's minimax\edit{$^\star$} & No &excellent & ok \\
	Zheng et al., 2011 & \cite{Zheng2011}& 3.2 a & mesh & Gaussian & Gaussian&good & good\\
	Zhang et al., 2015 & \cite{Zhang2015}& 3.2 b & mesh & Gaussian  & Gaussian&good & good \\
	Yadav et al., 2018 & \cite{Yadav2018}& 3.2 c & mesh & Tukey's  & Gaussian& excellent & good\\
	\"{O}ztireli, 2009 & \cite{2009Oztireli_FeatPresPCNonLinKernRegr}& 4.1 a & PSS & Gaussian & Gaussian&good & good\\
	Mattei and Castrodad, 2016 & \cite{2016Mattei_PCDenoisingMRPCA}& 4.1 b &PSS& Gaussian  & No&good & ok\\
	Li et al., 2009 & \cite{2009Li_FADenoisePSS} & 4.2 a & PSS & Gaussian & Gaussian&good & good\\
	Zheng et al., 2017 & \cite{2017Zheng_GuidedPCDenoising}& 4.2 b & PSS & Gaussian  & Gaussian&good & good \\
	Park et al., 2013 & \cite{2013Park_FAFilteringPSS}& 4.2 c & PSS& Gaussian  & No&good & ok\\
	Digne and Franchis, 2017 & \cite{2017Digne_BilateralFilterPC}& 4.2 d & PSS & Gaussian  & Gaussian&good & good\\
	Zheng et al., 2018 & \cite{2018Zhen_RollingNormalFilterPC}& 4.2 e & PSS & Gaussian  & Gaussian&good & good\\
	Yadav et al., 2018 & \cite{Yadav2018_VNVT}& 4.2 f & PSS & Truncated $L_2$ & No&good & ok
\end{tabular}
\caption{Overview on the discussed methods. For each method, we present the authors, year, citation, which input is processed (PSS or meshes), what error norm is used and whether a spatial weighting is applied. Furthermore, we collect the assessments from the above sections how the different methods perform in terms of feature-preservation and noise-removal.\\
\edit{$\star$ The error norm used in method~\cite{Centin2018} is not equivalent to Huber's minimax. However, the utilized weighting term closely resembles the function~$g_\sigma(x)$ of Huber's minimax, see Table~\ref{tab:estimators} and the discussion in \ref{sec:UnilateralNormalFiltering} g).}}
\label{tab:sota}
\end{sidewaystable}

\section{Conclusion}

In this paper, we unified state-of-the-art methods for normal filtering in surface denoising using the robust statistics framework. We discussed different \mbox{M-estimators}, which are the main tools of robust statistics. These tools are defined by a robust error norm and a corresponding influence function respectively. Based on the properties of the influence function (bounded and re-descending) and of the anisotropic weighting function, we discussed the robustness of state-of-the-art methods in terms of feature-preservation and feature-enhancement (see Table~2). Furthermore, we have shown that the introduction of spatial filters along with anisotropic filters will improve the robustness of the algorithm in terms of noise-removal. The robust statistics framework not only provides a platform to bring new insight into the field of surface-denoising and clarify the relation between different methods in the field. It can also be used for new methods to combine the advantages of the known filtering techniques. The application of robust statistics is not limited to surface denoising, it can be used effectively in other areas of the field of geometry processing. Corresponding applications of this powerful tool are left as further research.

\section*{Acknowledgments}

This research was supported by the DFG Collaborative Research Center TRR 109, ``Discretization in Geometry and Dynamics'', the Berlin Mathematical School, the Einstein Center for Mathematics Berlin, and the German National Academic Foundation. The authors would like to thank the anonymous reviewer for many helpful suggestions and comments on how to improve the article.

\bibliographystyle{unsrt}
\bibliography{article}
%
%
\end{document}